\documentclass[preprint]{aastex}
\newcommand{\het}{Hobby$^*$Eberly Telescope}
\shorttitle{Circumstellar Signatures in SN~2003du?}
\shortauthors{Gerardy et al.}
\begin{document}
\title{SN~2003du: Signatures of the Circumstellar Environment in a Normal Type~Ia Supernova?
\footnotemark[1]}
\author{
Christopher L. Gerardy\altaffilmark{2}, 
Peter H\"oflich\altaffilmark{3}, 
Robert A. Fesen\altaffilmark{4},
G. H. Marion\altaffilmark{3}, 
Ken'ichi Nomoto\altaffilmark{5},
Robert Quimby\altaffilmark{3}, 
Bradley E. Schaefer\altaffilmark{6},
Lifan Wang\altaffilmark{7},
J. Craig Wheeler\altaffilmark{3}
}
\footnotetext[1]{Based on observations obtained with the Hobby-Eberly Telescope, which is a
joint project of the University of Texas at Austin, the Pennsylvania State
University, Stanford University, Ludwig-Maximillians-Universitaet
Muenchen, and Georg-August-Universitaet Goettingen.}
\altaffiltext{2}{W.~J.~McDonald Postdoctoral Fellow, McDonald Observatory, University of Texas at Austin, Austin, TX 78712}
\altaffiltext{3}{Department of Astronomy, University of Texas at Austin, Austin, TX 78712}
\altaffiltext{4}{Department of Physics \& Astronomy, Dartmouth College, 6127 Wilder Laboratory, Hanover, NH 03755}
\altaffiltext{5}{Department of Astronomy, University of Tokyo, Bunkyo-ku, Tokyo 113-0033, Japan}
\altaffiltext{6}{Department of Physics and Astronomy, Louisiana State University, Baton Rouge, LA 70803}
\altaffiltext{7}{Lawrence Berkeley Laboratory 50-232, 1 Cyclotron Road, Berkeley, CA 94720}

\begin{abstract}

We present observations of the Type Ia supernova 2003du obtained with the \het\ (HET) and report 
the detection of a high-velocity component in the \ion{Ca}{2} infrared triplet near 8000
\AA, similar to features previously observed in SN~2000cx and SN~2001el.  This feature exhibits a 
large expansion velocity ($\approx$ 18,000 km~s$^{-1}$) which is nearly constant between $-7$ and 
$+2$ days relative to maximum light, and disappears shortly thereafter.  Other than this feature, 
the spectral evolution and light curve of SN~2003du resemble those of a normal SN~Ia. 

We consider a possible origin for this high-velocity \ion{Ca}{2} line in the context of a 
self-consistent spherical delayed-detonation model for the supernova.
We find that the \ion{Ca}{2} feature can be caused by a dense shell formed when circumstellar 
material of solar abundance is overrun by the rapidly expanding outermost layers of the SN ejecta.
Model calculations show that the optical and infrared spectra are remarkably 
unaffected by the circumstellar interaction and the resulting shell. 
In particular, no hydrogen lines are detectable in either absorption or emission after the phase of 
dynamic
interaction.  The only qualitatively different features in the model spectra are the strong, high velocity feature
 in the  \ion{Ca}{2} IR-triplet around 8,000~\AA,
 and a somewhat weaker \ion{O}{1} feature near 7,300~\AA.  
The Doppler shift and time evolution of these features provides an estimate for the 
amount of accumulated matter (decreasing Doppler shift with increasing shell mass) and also an indication
of the mixing within the dense shell.
For high shell masses ($\approx 5 \times 10^{-2} M_\odot$), the high-velocity component of the \ion{Ca}{2}\
line merges with the photospheric line forming a broad feature.  A cut-off of the blue wings of strong, 
un-blended lines (particularly the \ion{Si}{2} feature at about 6,000 \AA) may also be observable for larger
shell-masses.  The model SN~Ia light curves are little effected except at very early times when the shell is 
partially optically thick due to Thomson scattering, resulting in larger $(B - V)$ colors by up to $0.3^m$.

We apply these diagnostic tools to SN~2003du and infer that about $2 \times 10^{-2}M_\odot$ of 
solar abundance material may have accumulated in a shell prior to the observations. 
Furthermore, in this interpretation, the early light curve data imply that the circumstellar 
material was originally very close to the progenitor system, perhaps from an accretion disk, 
Roche lobe, or common envelope.  Because of 
the observed confinement of \ion{Ca}{2} in velocity space and the lack of ongoing interaction 
inferred from the light curve, the matter cannot be placed in the outer layers of the exploding 
white dwarf star or related to a recent period of high mass loss in the progenitor system prior to 
the explosion.  We note that the signatures of circumstellar interaction could be rather common in 
SNe~Ia and may have eluded discovery because optical spectra often do not extend 
significantly beyond 7500~\AA.

\end{abstract}
 

\section{Introduction}
 
There is general agreement that Type Ia Supernovae (SNe~Ia) result from some process involving the
combustion of a degenerate C/O white dwarf (WD) \citep{hf60}. Within this general picture, two
classes of models are most likely: (1) An explosion of a C/O-WD with a mass close to the 
Chandrasekhar limit ($M_{Ch}$), which accretes matter through Roche-lobe overflow from an evolved 
companion star \citep{Whelan73}. In this case, the explosion is triggered by compressional heating 
near the WD center.  Alternatively, (2) the SN could be an explosion of a rotating configuration 
formed from the merging of two low-mass WDs, after the loss of angular momentum due to gravitational
radiation \citep{webbink94,it84,pac85}.  Candidate progenitor systems have been observed for both 
scenarios: WD binary systems with the correct period to merge in a Hubble time and an appropriate 
total mass \citep{max2000}; and supersoft X-ray sources \citep{greiner91,vdh92,rap94,kah97} showing 
accretion onto the WD from an evolved companion. However, there are still open questions about the 
details of both the merging and accretion processes 
\citep[e.~g.][]{nomoto82,benz90,piersanti00,nomoto00,nomoto03}.

From the observed spectral and light curve properties, the first scenario appears to be the most 
likely candidate for the majority of normal SNe~Ia.  In particular, delayed detonation (DD) models 
\citep{k91,yamaoka92,ww94} have been found to reproduce the majority of the observed 
optical/infrared light curves (LC) and spectra of SNe~Ia reasonably well 
\citep{h95,fisher95,hk96,wheeler98,lentz01,h02}. In the DD scenario, a slow deflagration front turns
into a detonation.  We note, however, that detailed analyses of the observed spectra and light 
curves indicate that mergers may still contribute to the supernova population 
\citep{hk96,hatano2000}. For recent reviews see \citet{branch99,hn00,hetal03}.

In both of these scenarios, a certain amount of loosely bound material associated with the mass 
transfer is likely to remain in the system at the time of the explosion.  In the binary mass
transfer scenario, the donor star itself may be the source of such material.  Based on hydrodynamic 
calculations, \citet{mb00} found that the SN ejecta wraps around the donor star and, depending on 
the donor's evolutionary phase, may strip off up to several tenths of a solar mass of H/He rich gas.
As a result, H or He might be observed with expansion velocities of a few hundred km~s$^{-1}$. 
To date, however, no convincing evidence for this kind of stripped material has been observed. 

Indeed, a number of possible signatures of interaction with a circumstellar  
environment have been studied, including X-rays \citep{schlegel93,schlegel95}, radio \citep{bb95} 
and, most often, narrow absorption and emission lines due to \ion{H}{1}, \ion{He}{1}, and 
\ion{He}{2} lines \citep{chugai86,ltw92,branch95}.  However, most SNe~Ia show no 
observational evidence for any of these indicators. 
The most comprehensive search for \ion{H}{1} emission
lines been done for SN~1994D by \citet{cumming96} who found an upper limit for the progenitor mass 
loss of $1 \times 10^{-5} M_\odot$.

Similarly
in the case of WD mergers, the exploding star is expected to be surrounded by debris from the merging 
process which will not undergo thermonuclear burning \citep[e.~g.][]{benz90}.  \citet{k93} showed 
that the interaction of the supernova with this material may lead to a shell structure in the 
ejecta, and the observable consequences for the SN light curve and spectra have been studied 
subsequently \citep{hk96}.  Although a few events (e.~g. SN~1990N; \citealt{hk96}) show some 
evidence for this kind of structure, a strong case for such a shell has yet to be found.  

Thus, despite significant progress in our understanding of Type~Ia supernovae, we still have few 
observational constraints on the progenitor environment.
Likewise, little is known about the variety of the progenitor systems \cite{wheeler91}.  

However, the recent discovery of strong hydrogen lines in SN~2002ic \citep{hamuy2003} has drawn new 
interest in this subject, and emphasized the importance of detecting circumstellar material as a tool 
for understanding the progenitor system. Whereas most of the optical spectrum of 
SN~2003ic closely resembles that of a normal SN~Ia, it also exhibits H lines similar 
to those seen in SNe~IIn with both broad and narrow components.  Significantly more than 
$0.1 M_\odot$  of H-rich material is required to explain the features in SN~2003ic, with the H-rich 
gas at distances between $10^{16}$--$10^{17}$cm. This matter might be attributed to a short period
of high mass loss in a binary system or during a planetary nebula phase several thousand years 
before the explosion \citep{hamuy2003,wang2003b,livio2003}.

In this work, we revisit the question of circumstellar interaction in SNe~Ia in the context of a 
high-velocity component of the \ion{Ca}{2} infrared triplet feature, which we observe in the optical
spectra of SN~2003du.  Similar strong, high-velocity
\ion{Ca}{2} components have been observed in SN~2001el, an otherwise normal SN~Ia 
\citep{kris03}, and also in the unusual supernova SN~2000cx \citep{li01,thomas03}. In SN~2001el, this 
high-velocity
\ion{Ca}{2} feature was well separated in velocity space from the photospheric calcium, and was strongly 
polarized \citep{wang2003a}.  
\citeauthor{wang2003a}\ (\citeyear{wang2003a}; see also \citealt{kasen2003}) suggest that this 
feature in SN~2001el could be a consequence of nuclear burning in the WD 
(perhaps during the deflagration to detonation transition) which causes the ejection of a 
high-velocity, Ca-rich filament.  Alternatively, they suggest that it might be attributed to the 
surrounding accretion disk likely having undergone nuclear burning to increase the Ca abundance.

We note that observations of some other SNe~Ia have also shown a high-velocity component
of the \ion{Ca}{2} IR triplet (e.g. SN~1994D; \citealt{hatano99,fisher00}), which may be 
understood as a transient ionization effect when \ion{Ca}{3} recombines to \ion{Ca}{2} \citep{hwt98}.  
However, given the steep density gradient expected in the outer regions of the SN ejecta, it may be 
difficult to create a high-velocity \ion{Ca}{2} feature as persistent as those seen in SN~2000cx, 
and SN~2001el (or in SN~2003du).  

Through a quantitative
study of the formation of this IR Ca~II feature and related spectral properties, we examine a 
possible signature of solar abundance circumstellar matter.
In \S~2, we discuss 
the observations and data reduction. In \S~3, we use detailed 
models for the explosion, light curves and spectra to study the signatures of the accumulation of 
hydrogen-rich matter by a SNe~Ia. We then develop these signatures as diagnostic tools for probing
the circumstellar environment of SNe~Ia.  In \S~4, these tools are applied to the observations of 
SN~2003du, and we show that the spectral features are consistent with a H-rich shell formed 
by the interaction with matter related to the mass transfer in the progenitor system.  Finally, in
\S~5 we discuss the results in the general context and address the limits of our study.

\section{Observations and Data Reduction}
SN~2003du was discovered in UGC~9391 by LOTOSS \citep{schwartz03} on 22 April 2003 (UT) at about 
$15.9^m$. It was classified as Type~Ia by \citet{kotak03} on 24 April 2003 and 
resembled SN~2002bo about two weeks before maximum light.  

We obtained low-resolution (${\rm R} \approx 300$) optical spectra of SN~2003du 
using the Marcario Low Resolution Spectrograph (LRS; \citealt{hill98})
 on the Hobby$^*$Eberly Telescope (HET; \citealt{ramsey98}).
  For each epoch, two different setups  
were used giving effective wavelength ranges of 4100--7800 \AA\ and 5150--10~000
\AA.  The data for each setup were reduced separately using standard IRAF 
routines, and then flux-matched in the overlapping region and combined to
create a single 4100--10~000 \AA\ spectrum for each night.
Relative spectrophotometric calibration was accomplished by observing standards 
from \citet{massey88}, \citet{massey90}, and \citet{oke90}.  We are not able to 
obtain accurate absolute fluxes with LRS due to the time-variable effective 
aperture of the HET, and so the absolute flux levels of these spectra are only approximate.

\begin{figure*}
\includegraphics[width=10.7cm,angle=270]{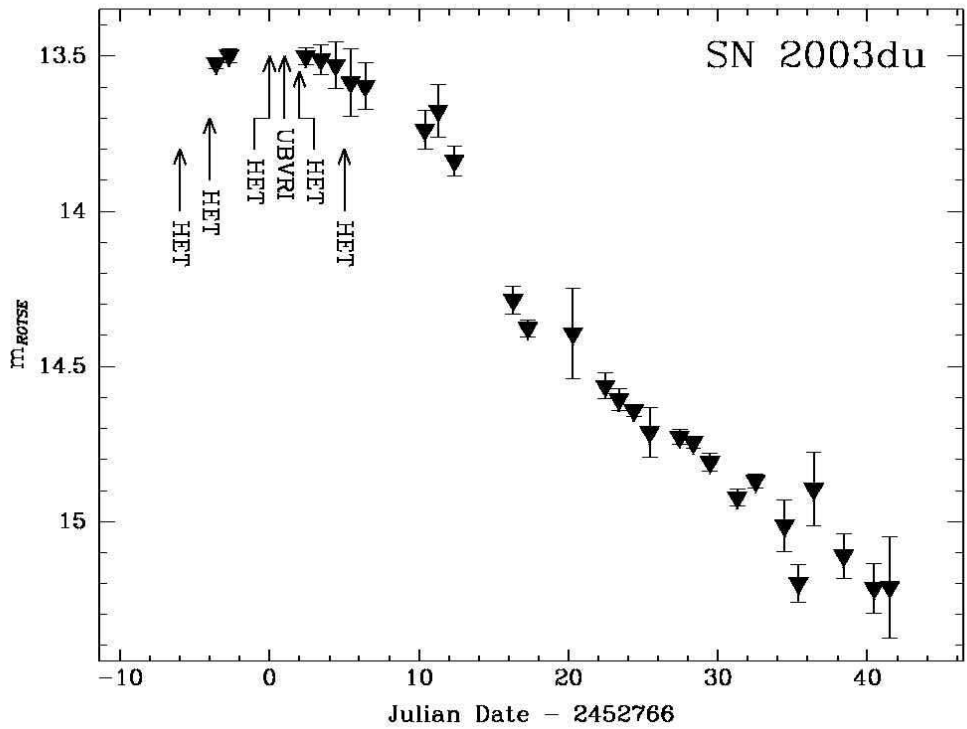}
\caption{
Unfiltered light curve of SN2003du as obtained by the automatic telescope
ROTSE. Dates at which HET spectra and UBVRI colors have been obtained are marked by arrows. 
From these data, SN2003du could be classified as a normal-bright SNe~Ia. Maximum light was at May 
07, 2003 with an uncertainty of $\pm 3 $ days.
}
\label{sn93du}
\end{figure*}

In addition to the HET spectroscopic observations, we also imaged SN~2003du 
with the 0.45~m robotic ROTSE~IIIb telescope \citep{akerlof03} to obtain an 
un-filtered broad-band light-curve.  ROTSE~IIIb observed SN~2003du at roughly 
1 hour intervals during each night when conditions were safe to open the 
telescope enclosure. 
The data were processed using the ROTSE automated 
reduction pipeline software, which delivers magnitudes for every source 
identified, relative to the red magnitudes of USNO A2.0 stars in the field. 
The pipeline photometry typically show a scatter of about 0.1 -- 0.2 mag for all
well-detected sources. (These data have been acquired during the commissioning phase
of ROTSE and test images suggest that the reduced CCD frames are significantly 
more accurate than the results of the current ROTSE pipeline.)
For this work, we averaged the
output of the pipeline data for each night and the resulting
light curve is presented in Figure~\ref{sn93du}.  The arrows indicate the timing of the
HET observations, as well as the timing of one epoch, near maximum-light, of 
{\it UBVRI} observations obtained using the 
0.8~m telescope at McDonald Observatory.

\begin{figure*}
 \includegraphics[width=13.7cm,angle=0]{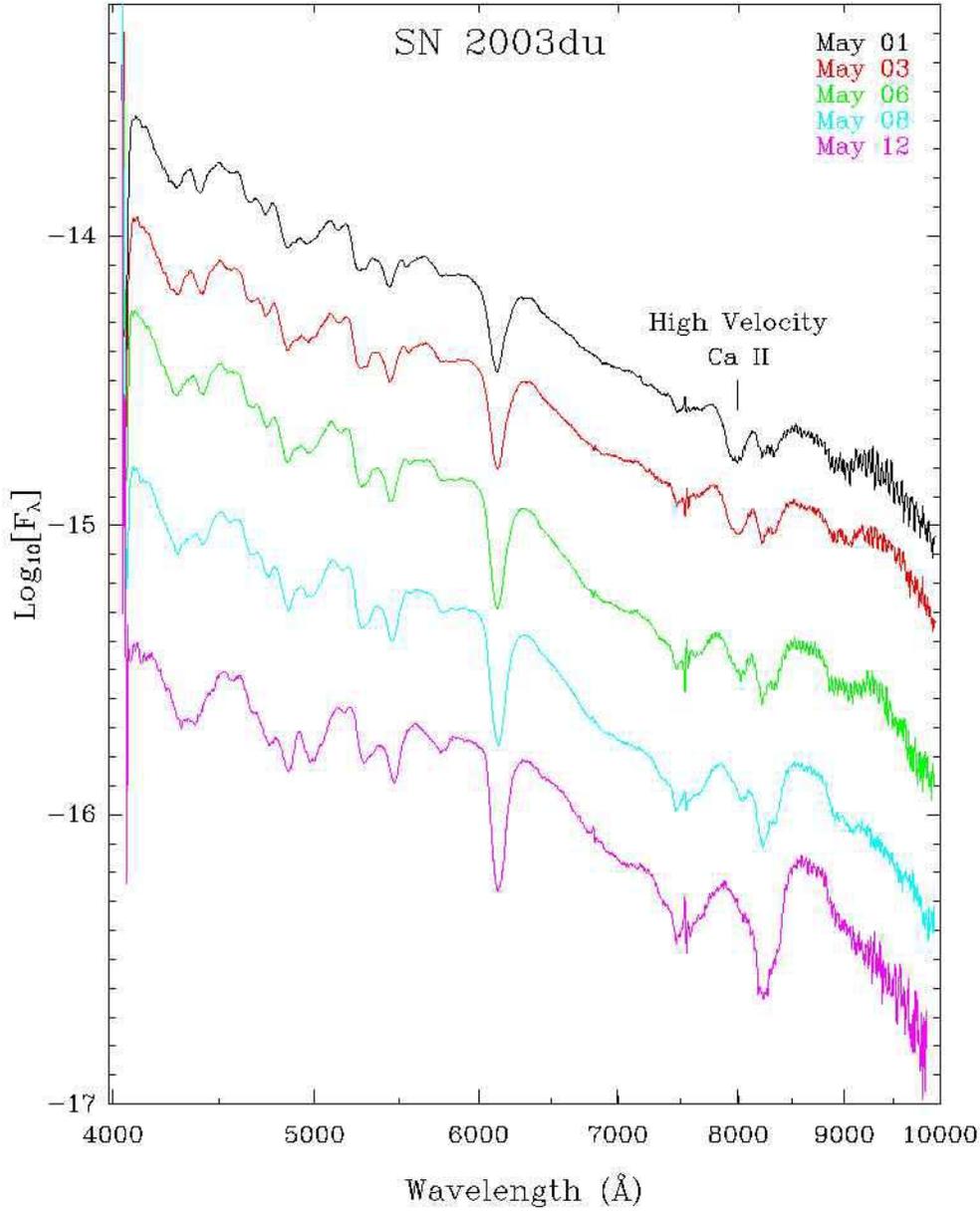}
\caption{Spectral evolution of SN2003du between -6 to +5 days relative to
maximum light in the ROTSE band. The high velocity component
of the \ion{Ca}{2} IR-triplet has been marked. The data have been shifted vertically by an
arbitrary amount for clarity.  The spectra are presented in the rest wavelength of the host
galaxy.}
\label{spectra}
\end{figure*}

\begin{figure*}
 \includegraphics[width=10.7cm,angle=270]{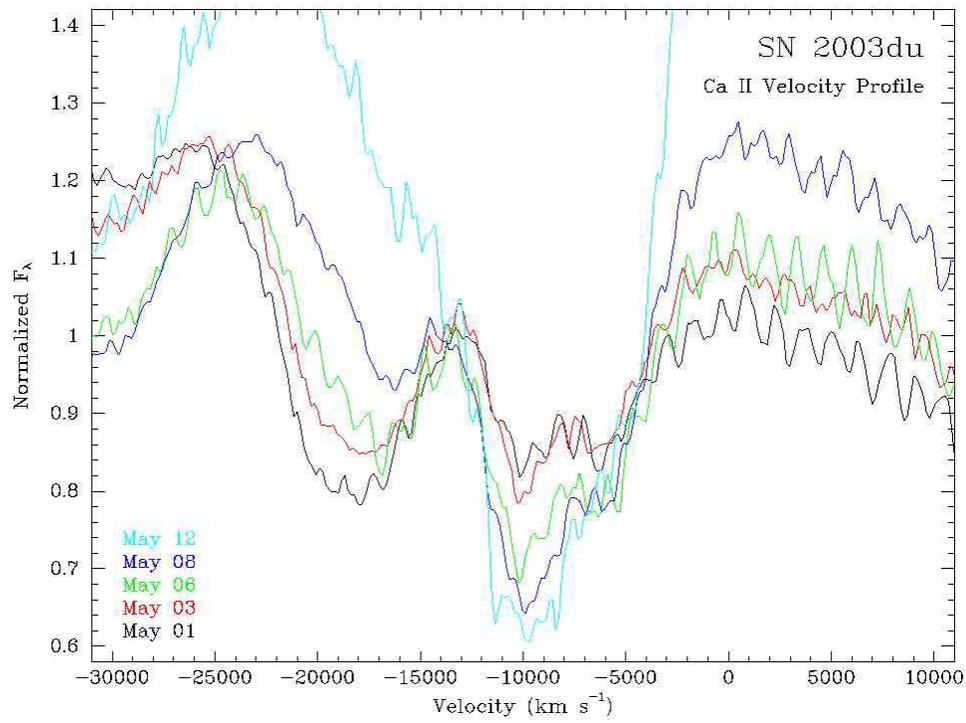}
\caption{Spectral evolution of the \ion{Ca}{2} IR feature between -6 and +5 days
relative to maximum light. The velocity scale is relative to the blue most line
at 8498 \AA. The gf-value weighted centroid of the IR triplet is at about +2,700 km~s$^{-1}$.
}
\label{spectra_ca}
\end{figure*}

The observed spectroscopic evolution of SN~2003du is presented in Figure~\ref{spectra}.  
These spectra exhibit a high-velocity blueshifted component in the \ion{Ca}{2} feature near 8000 \AA.
The Doppler shift of this feature ($\approx 18,000$ km~s$^{-1}$) remains 
well above the photospheric velocity ($\approx$ 11,000 km~s$^{-1}$) until the feature fades shortly
after maximum light.
Note that although the centroid of the absorption feature shifts somewhat to the red as it evolves
(Fig.~\ref{spectra_ca}), 
this is due to the blue side fading earlier than the red, rather than a shift of the entire feature.
The high-velocity absorption remains constrained in
velocity space, and in particular, the red edge of the absorption remains constant at about
-13,000 km~s$^{-1}$.  
In all other respects, the observed optical spectral evolution closely resembles that of a
normal Type~Ia supernova such as SN~1994D \citep{branch96,branch99}.
The near-maximum {\it UBVRI} colors (Table~\ref{colors}) of SN~2003du are also consistent 
with the expected colors of a normal SN~Ia \citep{phillips99}.

To provide further
constraint on the color evolution, we convolved the reduced HET spectra with 
the {\it BVRI} filter functions of \citet{bessel90} to obtain effective
photometry for each of the HET observations.  Zero-points for the effective
photometry were obtained by convolving the same filter functions with the 
synthetic spectrum of Vega presented by \citet{castelli94}. 
While the absolute fluxing of the HET data is poor, the relative 
spectrophotometry is accurate and thus the derived colors are meaningful. (Note that for the 
{\it B}-band we performed the convolution over only part of the Bessel bandpass, since the data
cut off below 4200 \AA.  For this reason, the HET $(B-V)$ colors are somewhat 
less accurate than the $(V-R)$ and $(R-I)$ colors.)
The resulting colors are presented in Table~\ref{colors},
along with the real {\it UBVRI} colors measured on 07 May 2003.  Comparison
of the HET derived colors on 06 May and 08 May with the {\it UBVRI} measurements 
suggest that the HET derived colors are probably accurate to about 0.1 mag.  When
compared to the \citet{riess96} templates, the observed SN~2003du colors 
lie within the ``1-$\sigma$'' region for normal SNe~Ia.

\section{Model Calculations}
To examine the effects of interaction with circumstellar material, we begin with a 1-D delayed
detonation model for the supernova.  The chemical and density structure of the outer regions of 
the SN ejecta are modified to model the hydrodynamic effect of the circumstellar interaction.
The result is then input into a radiation transport code to calculate
 synthetic non-LTE
 light curves and
spectra (see Appendix).

Previous analyses of the \ion{Ca}{2} feature for SN~2000cx and SN~2001el (\citealt{kasen2003,thomas03}) 
are based on 3D
parameterized density structures and assuming LTE population numbers, leaving a 
large number of free parameters. In contrast, we try to minimize the number 
of free parameters by using more realistic physics but a spherical configuration.
Our study presented here is based on detailed explosion models and NLTE light 
curves and spectra.
For example, in our analysis 
the velocity of the shell is linked directly
to the mass of the shell, and non-LTE provides a proper treatment of the ionization
balance and detailed atomic models,
needed since the level populations and opacities in supernovae are
are very different from LTE.  In the conclusions, we discuss the
limitations of our models and how asphericity will influence the results. 

\subsection{Delayed-Detonation SN model}
Our study is based on the delayed detonation scenario \citep{k91} which has been found
to reproduce the optical and infrared light curves and spectra of typical SNe~Ia reasonably
well \citep[e.g.][]{h95,hk96,nugent97,fisher95,wheeler98,lentz01,h02}. It also provides a natural
explanation for the brightness decline relation \citep{phillips93,hamuy96a,hamuy96b} as
a consequence of opacity effects in combination with nearly constant explosion energies for SNe~Ia 
\citep{h96,maeda03}.  

Hydrodynamic explosions, light curves, and synthetic spectra are all 
calculated self-consistently using only physically motivated connections between the different 
calculations.  Given an initial structure for the progenitor and a description of the nuclear 
burning front, the light curves and spectra are calculated directly from the explosion model 
without any additional tunable parameters.  This methodology forges a strong link between the 
physical processes being modeled and the predicted observables.  As a result, comparing the model 
results with observations can provide a great deal of insight, at the expense of having to perform 
rather more difficult calculations.

For this study, a single SN model (Fig.~\ref{model}) was chosen, which roughly matches the observed 
properties of 
normal Type~Ia supernovae.  However, no attempt has been made to fine-tune the SN model to 
``fit'' the observations of SN~2003du.  The details of the numerical methods and the resulting SN 
model used for these calculations are presented in Appendix A.  

\subsection{Circumstellar Interaction Model}
We consider the case of homologously expanding SN envelope running into a stationary circumstellar 
medium of solar-abundance gas. We model the region of circumstellar interaction in a manner similar 
to \citet{chevalier82} and the density profiles of the interaction region are based on his 
self-similar solution.  The collision of the SN ejecta and environment sets up a forward- 
and reverse-shock structure, separated by a contact discontinuity at a distance $R_c$.
This produces a high-density region with shocked ejecta and swept up matter, subsequently  
referred to as the shell, and a low-density precursor region of ambient gas.  The inner edge of the 
shell is given by the location ($R_{sh}$) of the reverse, adiabatic shock. Because the shock front is
Rayleigh-Taylor unstable \citep{chevalier82,dwarkadas98} we assume that the reverse shocked region 
consists of a mixture of SN and circumstellar matter. 
We use
a rather generic profile for shell because the details of the structure depend on
the exact location, origin and morphology of the environment. Moreover, we apply parameters,
which result in a  maximum density contrast of four between SN ejecta and the shocked region as 
expected for adiabatic shocks whereas other 
conditions, such as
nearby interaction (H\"oflich \& Khokhlov 1996)
or more sophisticated environments \citep{dwarkadas98} produce shells with higher contrast.
In such models, the strength of the individual features may be enhanced but we do not expect 
significant changes to our
deceleration
of the ejecta and the resulting position of the shell (in velocity) as a function of circumstellar mass.
These parameters depend on energy and momentum conservation and are thus relatively insensitive to such 
details.

For the calculation of synthetic spectra, 
we assume that the circumstellar matter originates close to the supernova and has been overrun 
by the SN ejecta well before the observations.  
As a consequence, the structure of the interaction region undergoes free homologous expansion in our
model structure, with little or no appreciable emission from the forward shock.
Furthermore, at the time of the observations, rapid expansion and radiative cooling will have erased the 
thermal signature from the shock interaction itself.
The temperature structure of the expanding shell is calculated
by taking into account radiative processes, gamma-ray heating, and adiabatic cooling. Thus 
the interaction will not significantly contribute to the luminosity if the interaction takes place
early on, i.e. at distances small compared to the SN photosphere at the time of the observation.
(As we discuss below, the light-curve of SN~2003du indeed suggests that any strong circumstellar 
interaction occurred before its discovery around two weeks before maximum light.)

We note that this model makes certain implicit simplifying assumptions, such as spherical symmetry,
power law density profiles for the ejecta and the surrounding medium, and pure adiabatic shocks.  
However, the qualitative results are relatively robust.  While deviations away from these
assumptions will tend 
to change the quantitative details of the resulting shell structure somewhat, the models are 
sufficient for the qualitative and order-of-magnitude quantitative analysis presented here.

\subsection{Model Results}

\subsubsection{General Considerations}
\begin{figure*}
 \includegraphics[width=12.7cm,angle=270]{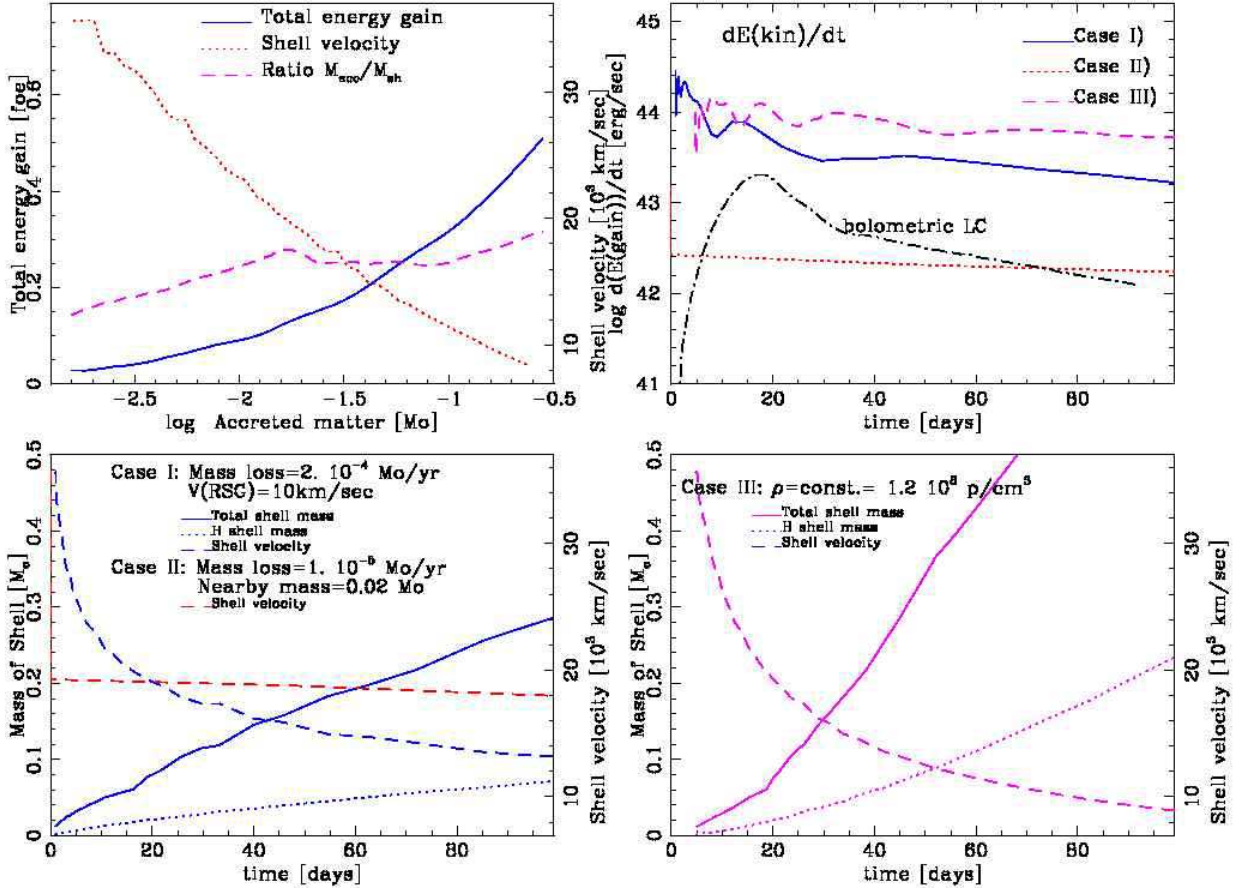}
\caption{Properties of the shell models.
 For the formation of the shell, we assume adiabatic shocks
and complete mixing during the interaction.
 The total energy gain by the interaction, the mean velocity of
the expanding shell, and the ratio between accumulated matter and the total shell mass
are given as a function of the accreted matter (above left).
Time dependent quantities are given for three cases
 which all produce a shell of $2\times 10^{-2}M_\odot $ at day 20.
 The shell is produced by running into a stellar wind with a velocity of 10 km~s$^{-1}$
(case I), a combination of a nearby mass and a small contribution due to stellar wind
 (case II)
, and an environment
with constant density. In the upper right, we give the energy generation rate
in comparison with the bolometric LC of the model without interaction. In the lower plots, 
we give the
the total mass of the shell and its velocity.
For details, see \S~3.
}
\label{shell}
\end{figure*}

The basic result of the interaction is the accumulation of circumstellar matter (of mass $M_{acc}$) 
in a shell (of mass $M_{sh}$). As the highest velocity ejecta runs into the reverse shock, it is 
decelerated down to the shell velocity $v_{sh}$.
Some of the basic quantities of the shell are shown in Figure~\ref{shell}.
Both the shell velocity and the total amount of kinetic energy converted by the interaction 
depend on $M_{acc}$. Typically, the shell consists of about 1/3 accumulated solar-abundance 
circumstellar matter and 2/3 SN ejecta with a composition depending on the shock velocity 
(compare Figs.~\ref{model} \& \ref{shell}). 

For a given SN explosion model, the relation between the shell mass and its velocity is fixed. 
Within the delayed-detonation scenario, the energy generation is nearly independent of changes in 
the explosion model because for $M_{Ch}$ models, the mass and density structure of the SN 
envelope hardly changes, although the chemistry of the outer envelope is different \citep{h02}. 
Conversely, only part of the C/O WD is burned in a pure deflagration scenario, reducing the 
kinetic energy of the ejecta. As a consequence, we expect lower shell velocities for those models.

As example cases, we consider shells which are produced by running into (case I) a stellar wind 
with a velocity $v_{RSG}$, (case II) a combination of a nearby mass and a stellar wind, or 
(case III) a constant density environment. For the first two cases, the mass accumulated in the 
shell at time 
$t$ can be obtained by scaling the relation for a constant mass loss rate
$$M_{acc}(t)= M_{acc}(t=0)+ \dot{M} {v_{sh}(t) \over v_{RSG}} \times t $$
where $v_{sh}$, $v_{RSG}$ and $\dot{M} $ are the shell velocity, the wind velocity of the progenitor
system and the mass loss rate, respectively. For an environment of constant particle 
density $N$ (case III), we have $$M_{acc} = {4 \pi \over 3 N_{av}} \mu _e * N \times R_c^3$$
where $\mu _e$ and $N_{av}$ are the mean molecule mass and Avogadro's number, respectively.

For this discussion, we consider H-rich shells produced by the accumulation of
$M_{acc} (t=20d) = 2.\times 10^{-2}$M$_\odot$. The properties of the system are
 $\dot M = 2.\times 10^{-4} $M$_\odot$~yr$^{-1}$ and $v_{RSG}=10$ km~s$^{-1}$ (case I),
 $\dot M = 1.\times 10^{-5} $M$_\odot$~yr$^{-1}$ plus an early accumulation of $1.98 \times 10^{-2} 
 $M$_\odot$,
 (case II), and a constant particle density $N=1.2 \times 10^{8}$g~cm$^{-3}$
 (case III).

\subsubsection{Conversion of Kinetic Energy}

 A general study of light curves for interactions between the SN shell and the environment
is well beyond the scope of this paper. Therefore,  we want to present a qualitative discussion
which let us to select our setup for the detailed calculation. The arguments are based on
energy and  momentum conservation and should apply rather generally but, as a consequence,
the conclusions are rather qualitative, and should be regarded as a check for consistency rather
than than quantitative.

In the upper right of Figure~\ref{shell}, we plot the rate of kinetic energy conversion in the 
shock. Though some of this energy will go into turbulent motion, 
and ionization of the gas, 
we expect significant modification of the
X-Ray/UV/optical radiation \citep{fransson96}. To decelerate the highest velocity ejecta to about
about 20,000 km~s$^{-1}$ as observed in SN2003du,
the energy gain by continuous accretion exceeds those of the
 underlying SN light curve (powered by radioactive heating via $^{56}$Ni and $^{56}$Co  decay)
by an order of magnitude.
 Thus, we expect significant modification of the light curves for 
case I and III, and even case II unless the wind component is very small. 

For SNe~Ia, such as SN~2003du, that exhibit essentially normal light curves, we can therefore rule 
out that the accumulated matter originates
predominantly
 from mass loss over an extended period of time or from a 
constant density environment.  In our light curve (Fig.~\ref{sn93du}), we do not see any 
significant additional energy input 60 days after the explosion (about 40 days after maximum) which,
in the context of this analysis,
limits  $\dot M$ to about $ \leq 1. \times 10^{-5} $M$_\odot$~yr$^{-1}$.
 We note that this limit is
consistent with upper limits for the mass loss in SN~1994D by Cumming et al. (1996) based on the 
lack of observed $H_\alpha $ emission.
However, since we have not calculated detailed self-consistent LCs with ongoing interaction it 
is not clear exactly how much ongoing interaction could be actually hidden in normal SNe~Ia.
More detailed analysis may require correction factors for this simple analysis
although LCs and spectra of SNe are rather sensitive to a change in the energetics.
 In any case, it is worth noting that the additional energy will likely dominate the LC at some
point in time and decrease its slope. However,
late time light curves of typical SNe~Ia do not show evidence for any additional energy 
source even after several years \citep[e.~g.][]{wells94,schmidt94}.

\subsubsection{Light Curve and Spectral Signatures}
In light of this constraint, we will consider shells that are formed early on in absence of continuous 
mass loss (case II with $\dot M = 0$).  Such a shell might originate, for example, from an 
accretion disk, the Roche-lobe of a companion donor star, or a common envelope of the progenitor 
system.  To construct the density 
structure of the expanding shell, we assume the 
ejecta freely expand out to $10^{13}cm$ before running into circumstellar matter of constant 
density with a thickness of $R_C-R_{sh}$.  To construct the shell, we assume power law densities 
for the outer SN ejecta ($\rho_{SN} \propto r^{-n}$), with $n$ given by the explosion model (close to 
$n \approx 7$; see Fig.~\ref{model}) 

\begin{figure*}
 \includegraphics[scale=0.75,angle=270]{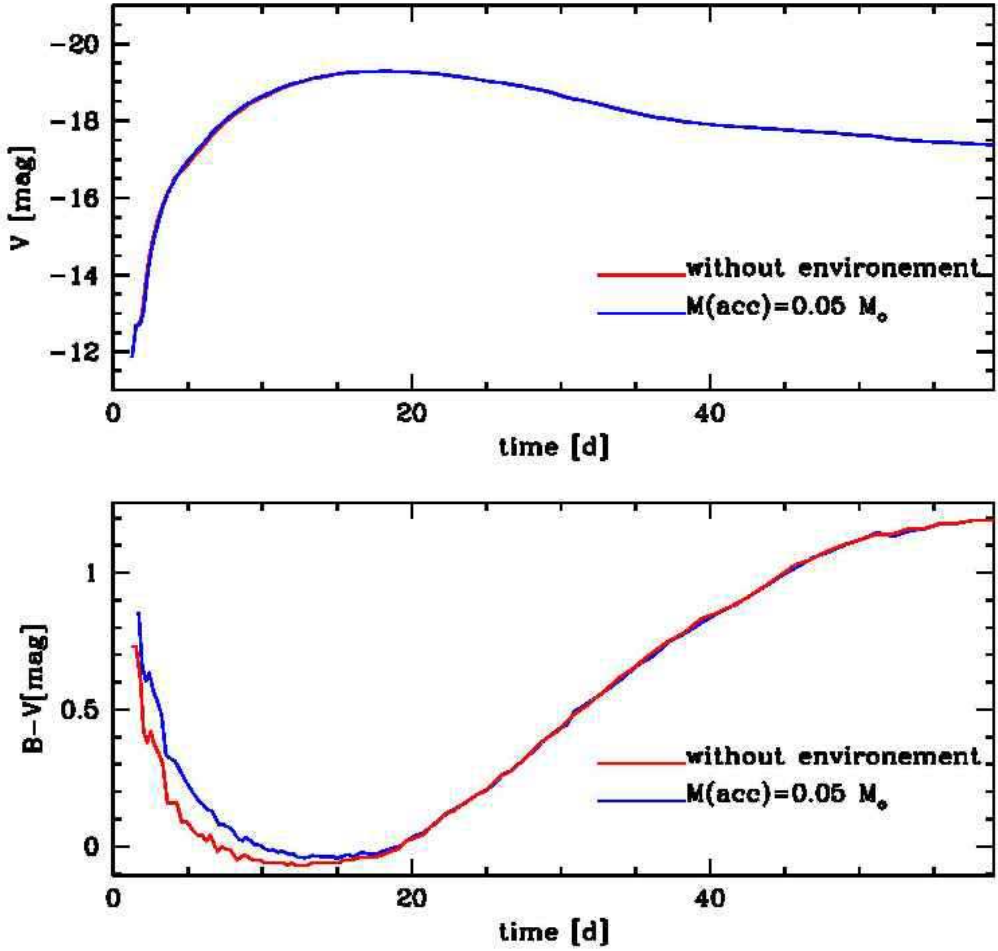}
\caption{Influence of an accumulated shell on the visual light curve and the color index B-V.
The light curves are shown for models  without a shell in comparison with H-rich, solar metallicity shells
of 5$\times 10^{-2}M_\odot$.  
Both models are plotted in the upper panel, but the \textit{V} light curves differ
by less than 0.1 mag (about the level of the internal accuracy of the models) and are virtually 
impossible to distinguish. 
Note that our observed LC starts later  (Fig.  \ref{sn93du}) and we can only conclude
that the observed LC is consistent with a normal SNe~Ia.
}
\label{model_lc}
\end{figure*}

\begin{figure*}
 \includegraphics[width=12.7cm,angle=270]{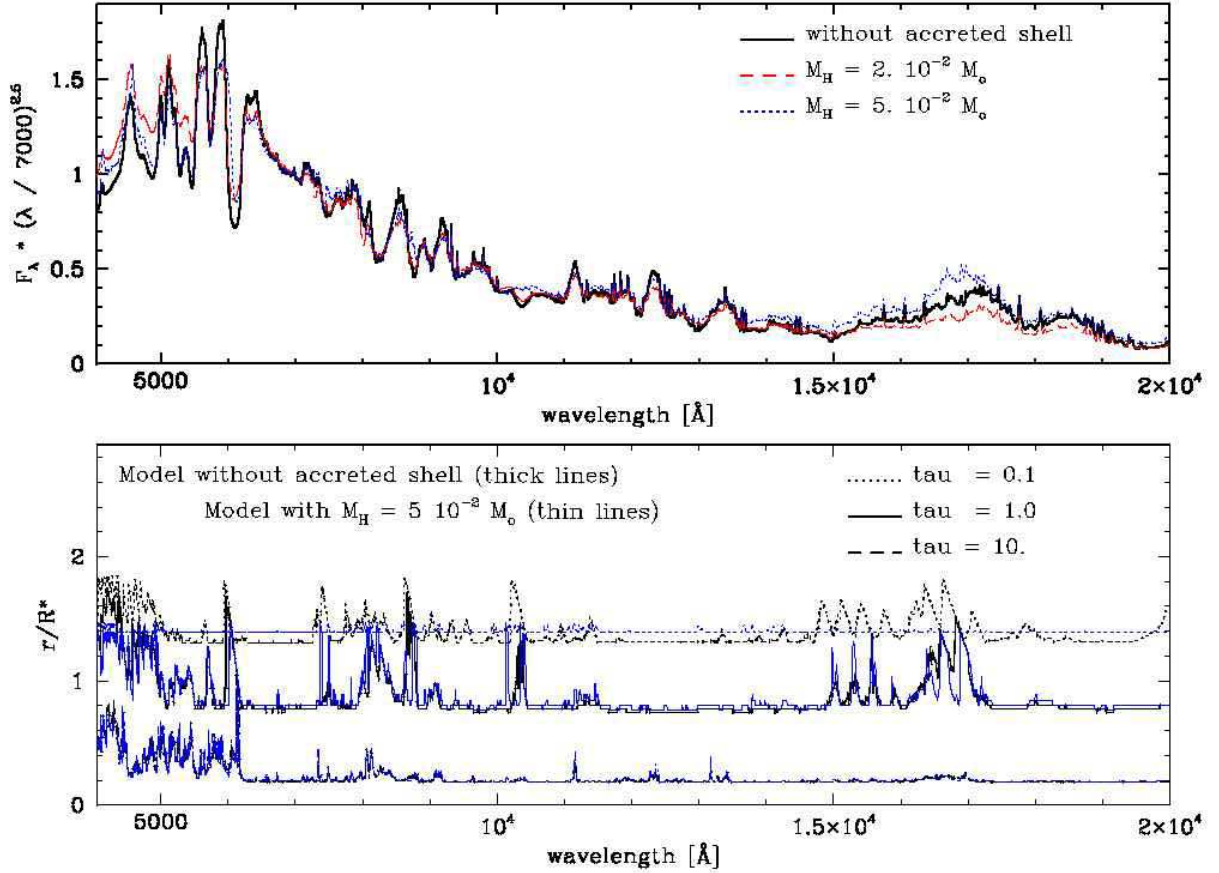}
\caption{Influence of an accreted shell on the synthetic spectra (normalized at 7000 \AA)
 between  3,800 and 15,000 \AA\ at $-3$d relative to $V_{max}$. Spectra are shown for a model 
without a shell and models with H-rich, solar metallicity shells of $2$ and
5 $\times 10^{-2}M_\odot$. Overall, these shells have little effect on the spectrum although 
continuum  scattering and backheating causes a ``smearing-out'' of line profiles.
The reduced expansion velocity of the outer layers results in a cutoff of the blue wings
in the absorption components of strong lines.
 We plot $F_\lambda \times ({\lambda \over 7000 \AA })^{2.5}$
to show simultaneously both the emission and absorption features in the optical and IR.
 The rather small influence of the shell can be understood by its small optical depth which
hardly changes the optical depth and physical conditions in the region where the photospheric
spectrum of the SN is formed.
 In the lower plot, we give the radius (normalized to $R_\star =10^{15}cm$) at which the 
 monochromatic Sobolev optical depth equals -0.1, 1 and 10, respectively,
 for the model without a shell (black) and $M_{acc}=5 \times 10^{-2}$ (blue).
}
\label{model_spec}
\end{figure*}

Light curves and detailed spectra have been calculated for a SN~Ia without a shell, and for cases with 
shell masses of 
$2,$ and $ 5 \times 10^{-2} M_\odot $ at several epochs.
 In Figure~\ref{model_lc}, the evolution of $V$ and $(B-V)$ is given. 
 Overall, even the high mass shell has comparably little effect on the
light curves because it hardly influences the diffusion time scales and the lack of energy generation without
ongoing interaction (see Figs. \ref{shell} \&  \ref{model_spec}). Because the Thomson optical depth of the shell
scales with about $t^{-2}$, it is partially optically thick up to about 10 days after the explosion.
As a result, the model photosphere becomes somewhat cooler and redder ($\Delta (B-V) \leq  0.3^m$) but 
still within the observed variations for normal SNe~Ia.

For illustration, optical and near IR spectra are shown in 
Figure~\ref{model_spec} for day 15 (about 3 days before maximum light).
We chose to exhibit this epoch of the spectral evolution as an example because the high-velocity 
\ion{Ca}{2} feature is well developed (the photosphere is far inside the shell region). Furthermore, 
at this epoch, the transient high-velocity feature due to the recombination of \ion{Ca}{3} to \ion{Ca}{2} 
is at its most pronounced \citep{hwt98}.  Thus thus epoch represents a ``worst-case'' for distinguishing 
between the shell and ionization interpretations of the high-velocity calcium feature.  We will discuss
this potential complication in detail in \S~\ref{HVCa_section}.

Overall, the model spectra with the three different shell masses are remarkably similar, although
large electron cross sections and backscattering cause some ``smearing out'' of the line features in
the higher shell mass models.  This 
insensitivity of the overall spectrum to the shell mass is a result of the large distance between 
the shell and the line forming region.  Nearly all of the spectral features are formed in regions 
well inside the shell, and are therefore largely unaffected by the interaction.  
There are a few places 
where \ion{O}{1} and \ion{Ti}{2} lines form subtly different features in the model spectra.  However,
these are highly blended with lines of singly and doubly ionized iron-peak elements.  Furthermore,
these weak features are below the accuracy of our numerical models (primarily due to the depth
discretization; see Appendix).  Models with an analytic density structure are more suitable for
investigation of these features.  (See Branch et al.~2003).

\begin{figure*}
 \includegraphics[width=12.7cm,angle=270]{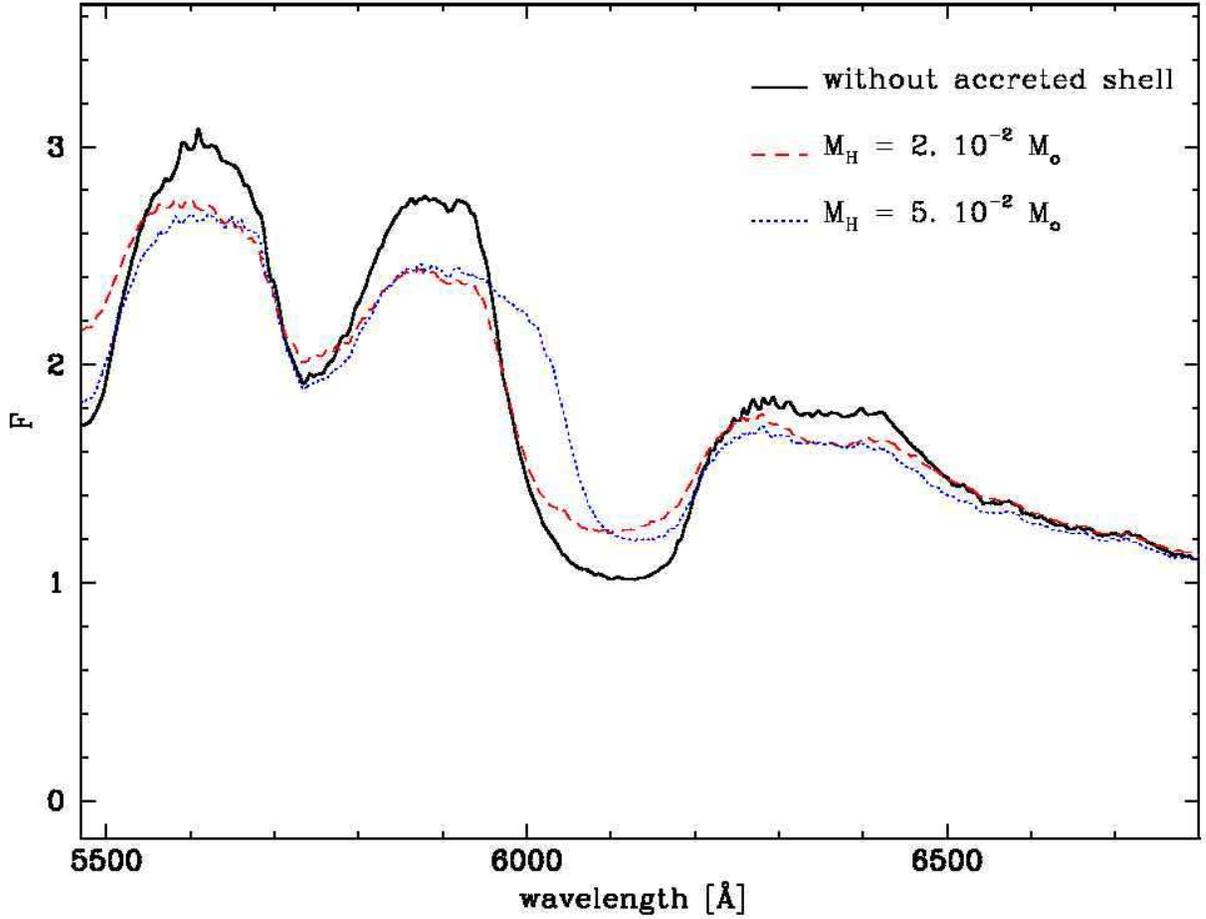}
\caption{Same as Figure \ref{model_spec} with the spectrum enlarged between 5500 and 6740 \AA.
 Changes become significant for a shell mass with $5\times 10^{-2} M_\odot$ but
are marginal for $2\times 10^{-2} M_\odot$. The main effects of a larger mass are a cutoff of the
blue edge of Si and a ``smearing'' out of features by continuum  scattering.
 No hydrogen lines can be seen (see \S 3.2.2). The small scale noise below 
the spectral resolution limit ($R \approx 1000$) is a measure of the internal accuracy of the models (see \S 3.1).
}
\label{model_si}
\end{figure*}

In most optical and IR features, 
the main difference is the cutoff at the blue edge of strong lines due to the 
deceleration of the high-velocity ejecta by the reverse shock (see Fig. \ref{model_si}).
Still, for the most part, the shell has little effect on the spectra beyond quantitative changes 
at the level of the intrinsic variability in normal SNe~Ia.  Even shells with about 
$5 \times 10 ^{-2}M_\odot$ might elude discovery.  

However, there are two features that exhibit clear qualitative rather than quantitative changes with 
increasing shell mass.  In the presence of a shell, high-velocity components appear in the 
\ion{Ca}{2} IR triplet at about 8000 \AA\ and in a feature between 7300 and 7500 \AA\ due to a blend
of \ion{O}{1}/\ion{Mg}{2} (Fig.~\ref{model_ca}).  Because these features are qualitatively 
different in the interaction scenario, they can be used to probe for such an interaction without
having to rely on the detailed quantitative accuracy of model spectra.  

With a strong high-velocity feature forming in the \ion{Ca}{2} IR-triplet, a similar feature may be expected in
the \ion{Ca}{2} H\&K lines.  However, in our models the spectral region below 4000 \AA, containing
the \ion{Ca}{2} H\&K lines, is highly contaminated with other features. At early times lines of Fe/Co/Ni III 
populate this spectral region, and beginning about 7 days before maximum light a strong \ion{Si}{2} 
appears at the expected wavelength of a high-velocity component for the \ion{Ca}{2} H\&K features.  In contrast,
the \ion{Ca}{2} IR triplet lies in a comparatively empty spectral region, making it much better 
suited as a spectral diagnostic.  For further discussion of high-velocity components in the \ion{Ca}{2} H\&K
lines, see Branch et al.~2003.

\subsubsection{High-Velocity \ion{Ca}{2} IR Triplet Feature\label{HVCa_section}}

\begin{figure*}
\includegraphics[width=8.7cm,angle=270]{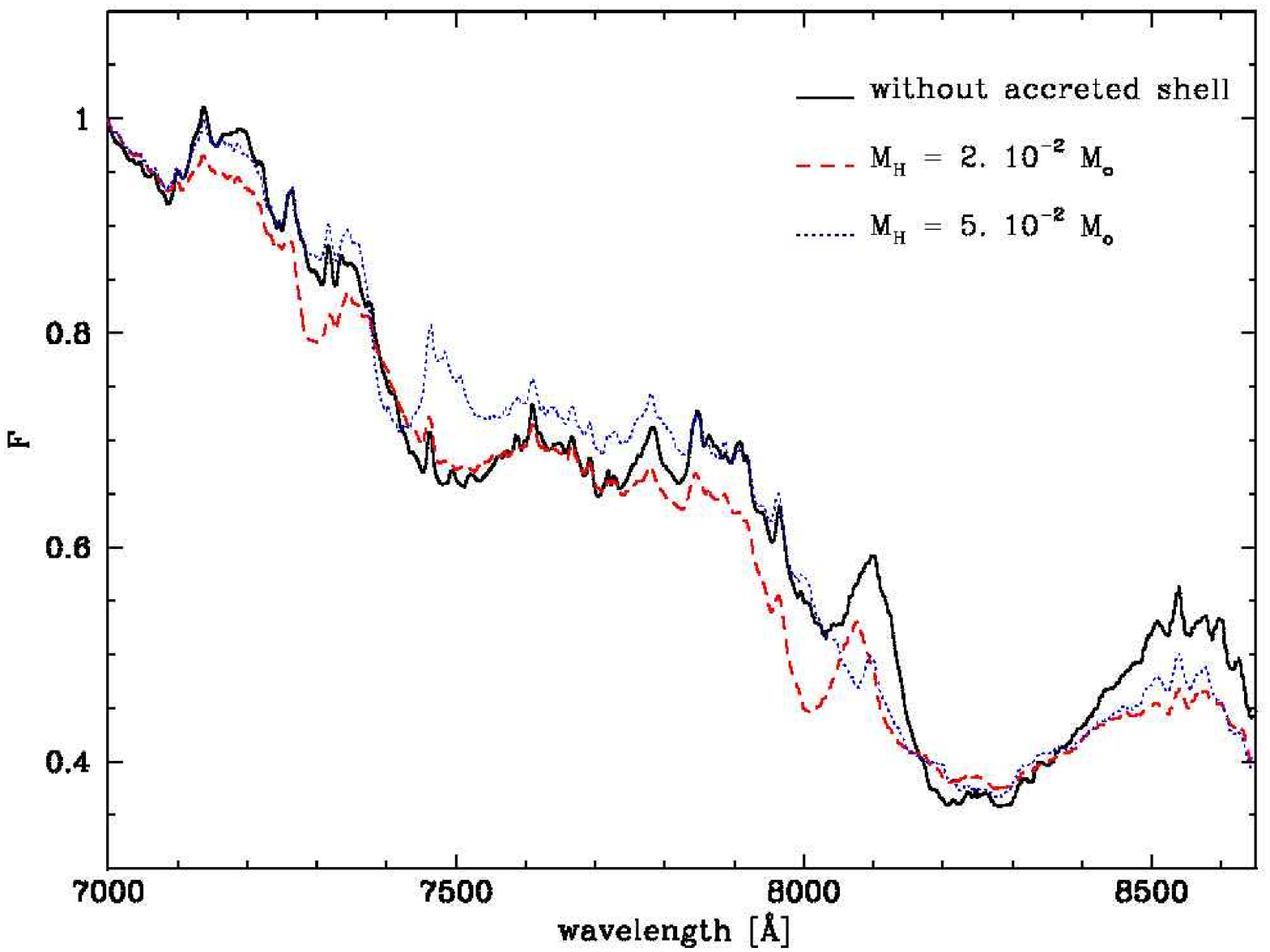}
\includegraphics[width=8.7cm,angle=270]{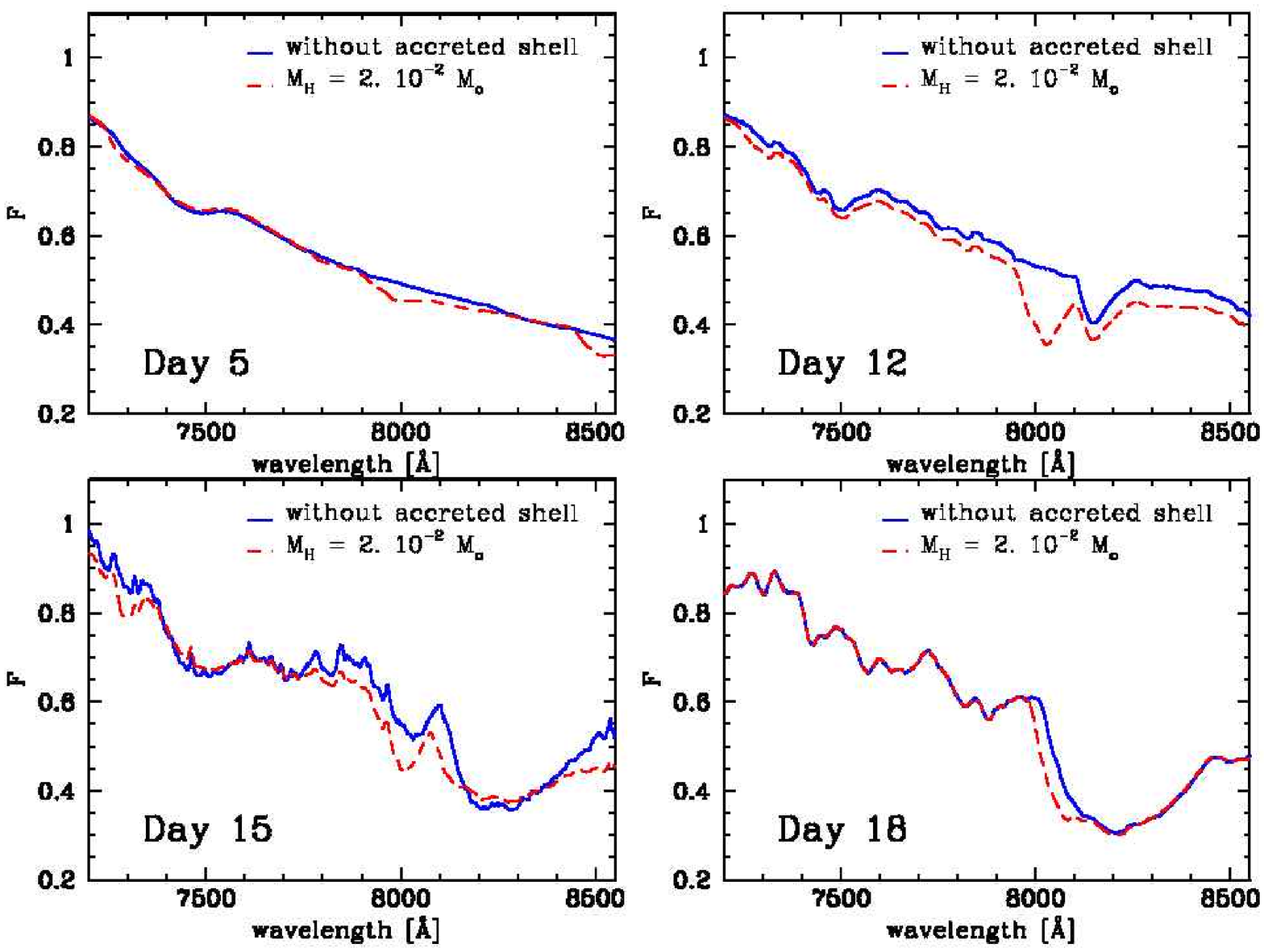}
\caption{Same as Figure \ref{model_spec} between 7000 and 8600 \AA\ which is the only
region with qualitative changes of individual lines even for the model with $M_H$ of
$2 \times 10^{-2}M_\odot$. Features include an OI/MgII blend
at 7300 \AA\ and the high velocity \ion{Ca}{2} absorption at about 8000 \AA .
 In the upper plot, a detailed comparison is given for all models at day 15 after the 
explosion. In the lower plots, we show the spectral evolution with time between day 5
to day 18 for the model without and a shell with $2 \times 10^{-2} M_\odot$.
 The dominant signature of interaction is the appearance of a secondary, high velocity
Ca II feature or, for high shell masses, a persistent high velocity component in a broad
Ca II line which starts to appear at about 5 days after the explosion.
 Note that, even without a shell, a secondary Ca II feature can be seen for a
period of 2 to 3 days during the phase when Ca III recombines to Ca II, emphasizing the
importance of a good time coverage for observations.
}
\label{model_ca}
\end{figure*}

In the models which include a shell from circumstellar interaction, a rather strong high-velocity 
component appears in the \ion{Ca}{2} IR triplet (Fig.~\ref{model_ca}), with a Doppler shift corresponding 
to the expansion velocity of the shell (Fig.~\ref{shell}) rather than the photospheric velocity. In 
combination with explosion models (but actually relatively insensitive to the model details within a 
delayed-detonation scenario), this provides a good measure for the mass of circumstellar material 
accumulated in the shell. 
For example, with $M_{acc} = 2 \times 10^{-2}M_\odot$, and $ 5 \times 10^{-2}M_\odot$ the 
high-velocity \ion{Ca}{2} Doppler shift corresponds to about 19,000 km~s$^{-1}$ and 14,000 
km~s$^{-1}$, respectively.  In the model with the higher shell mass the
deceleration of the outermost SN ejecta is sufficient to cause a partial merging of the high velocity 
component with the photospheric line.  The resulting feature looks more like a very broad feature rather 
than two distinct components.

However, Figure~\ref{model_ca}\ also illustrates a complication such an analysis. At the epoch 
shown in the top panel of Figure~\ref{model_ca}\ (about three days before maximum light), the model 
without an accreted shell also shows a strong high-velocity component to the IR triplet feature.
As we noted in the introduction, observations of some SNe~Ia (e.g. SN~1994D) show a short-lived 
high-velocity component of \ion{Ca}{2} IR triplet shortly before maximum light. In SNe~Ia where the 
high-velocity \ion{Ca}{2} feature is short-lived, or observed at only a single epoch near maximum light, 
this feature can also be interpreted as a transient ionization effect as \ion{Ca}{3} recombines to 
\ion{Ca}{2} \citep{h95,hwt98,lentz01}.  During this recombination phase, models for normal SNe~Ia 
show an outer and inner region of \ion{Ca}{2} separated by \ion{Ca}{3}. The recombination from 
\ion{Ca}{3} to \ion{Ca}{2} occurs around maximum light over the course of a couple of days.
As a result, a two-component \ion{Ca}{2} feature due only to this recombination effect is rather
short-lived, and the Doppler shift of the high-velocity 
absorption component recedes rapidly and merges with the photospheric \ion{Ca}{2} feature.  
Because of the steeply declining density gradient in the ejecta, a high-velocity component formed in 
this way is rather weak, only about 20\% relative to the continuum in our model.  

In contrast to this transient ionization effect, the high-velocity component formed by a dense shell
evolves much more slowly, and has a Doppler shift that is nearly constant if the mass of the shell is 
not significantly changing.  The time evolution of features due to these two effects are compared in the 
lower panels of Figure~\ref{model_ca}.  The high-velocity feature in the shell model can be seen as 
early as 5 days after the explosion, while the feature due to the ionization effect is visible only in
the day 15 spectrum.  In both cases, the feature disappears around maximum light.

Note that even in a SN with a statically evolving shell, a slight shift in the centroid of the 
high-velocity absorption is to be expected. Since the shell is
also undergoing free homologous expansion, the outer regions of the shell will undergo geometric
dilution faster than the inner edge.  As a result, the feature fades faster
at the blue, high-velocity end than the red end, shifting the feature's centroid to the red.

\subsubsection{High-Velocity \ion{O}{1} Feature}
The other new feature in the models which include a shell 
is an absorption between 7300 and 7400 \AA\ due to an \ion{O}{1} blend
(7775, 7774 \& 7772 \AA).  This feature shows the same Doppler shift and evolution 
with time as the \ion{Ca}{2} feature (i.e.~related to the dynamics of the shell and not
the photosphere).  However, unlike the Ca feature, the oxygen is not primordial but produced 
during explosion. Burned matter contributes about 2/3 of the shell mass (see Figs. \ref{model} \& 
\ref{shell}). Solar abundances of oxygen do not provide sufficient optical depth in the shell to
cause a strong feature. The appearance of this feature would imply that a significant amount of 
burnt SN ejecta has piled up in the dense shell.

\subsubsection{Hydrogen, Helium, \& Carbon}
Although 20 \% of the shell consists of hydrogen, the model spectra do not exhibit any significant 
Balmer or Paschen lines, even for $M_{acc} $ of $5 \times 10^{-2} M_\odot$.  In the 
absence of heating by an ongoing interaction between the SN and its environment, the temperature in 
the shell is low ($T\approx$ 4500 -- 5000~K).  At these temperatures, excitation of metal lines is
strongly favored over hydrogen, due to the much lower excitation energies. (e.g.\ 10.2~eV for 
\ion{H}{1} vs.\ 2.22~eV ($1D \rightarrow 1S$), 9.0~eV ($3P \rightarrow 3S$), 1.57~eV ($5S 
\rightarrow 5P$) for the singlet, triplet, and quintet state of calcium and, 1.9~eV for 
collisional coupling between the singlet and triplet state).  The optical depth of the shell is 
small for $\gamma $ rays ($\leq 0.01$ in all cases), keeping the non-thermal excitation of 
hydrogen low, and thus the hydrogen is mainly neutral.  Since most \ion{H}{1} atoms are in the 
ground state, the optical depth of the shell is very small for Balmer and Paschen lines and no 
absorption features are produced in the models.  
Furthermore, the models do not predict any significant emission component due to hydrogen. In the 
presence of heavy elements, strong charge exchange reactions between hydrogen and metals (with a 
lower ionization potential) are the preferred path for H recombination rather than the radiative 
process.  These same arguments hold even more true for \ion{He}{1}.
Note that while strong charge exchange reactions are expected to suppress H and He features, they 
are not really needed because, even without microscopic mixing,
detection of the broad, weak H emission features ($\approx$ 1 to 2 \%  of continuum in case II)
would hardly be possible.

On the other hand, because carbon and oxygen have excitation energies and line cross sections closer
to those of Ca, we could expect strong \ion{C}{1} features due to transitions at 9405~\AA, and 
10691~\AA\ (as was observed in the subluminous SN~1999by; \citealt{h02}). In normal-bright delayed 
detonation models, 
these \ion{C}{1} lines do not show up strongly because nearly all of the WD is incinerated, leaving
little carbon in the ejecta. However, in alternative scenarios for SNe~Ia such as deflagration 
and merger models, more than 0.1 $M_\odot$ of the WD remains unburned. For example, in the W7 model
\citep{nomoto84}, about $0.17 M_\odot$ does not undergo nuclear burning and, as a consequence, 
approximately equal amounts of O and C would be seen in the shell.  
A clear detection of this high-velocity \ion{O}{1} feature in would indicate that the outermost ejecta
had been trapped by the circumstellar interaction and piled up in a dense shell.  Any significant ammount 
of unburned carbon in these outer layers would also necessarily result in strong high-velocity features of 
\ion{C}{1}.  A lack of high-velocity \ion{C}{1} in the presence of high-velocity \ion{O}{1} would therefore
argue for nearly complete burning of the WD progenitor.

\section {Comparison with Observations}
\begin{figure*}
\includegraphics[scale=0.65,angle=270]{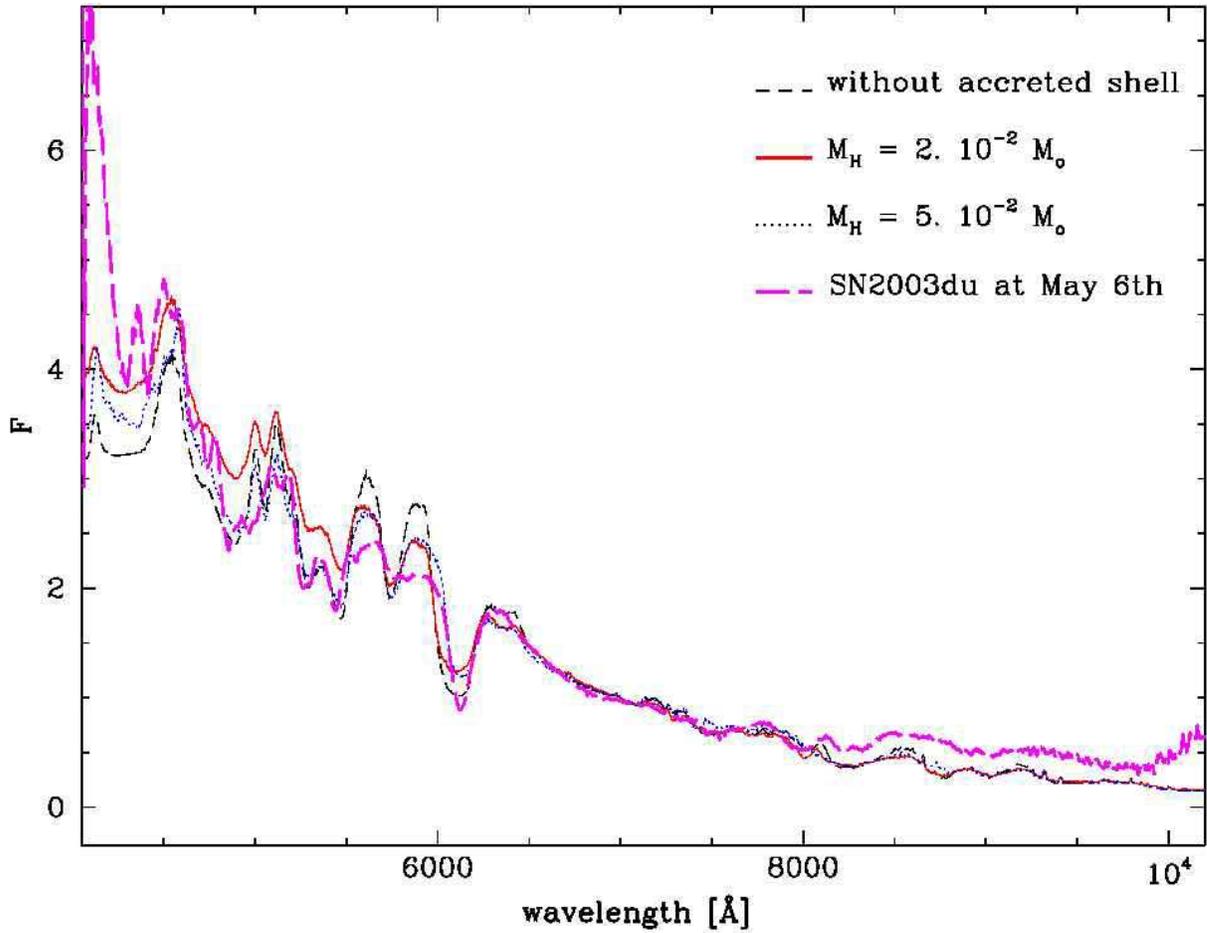}
\caption{Comparison of theoretical model at day -3 and SN~2003du on May 6th. 
The slope change at about 8,000 \AA\ is an artifact related to the 
normalization of the red and blue data and uncertainties in the instrumental response correction 
in the overlapping region.  Beyond 10~000 \AA\, the spectrum is contaminated by order overlap. 
}
\label{obsthe}
\end{figure*}

What follows is a comparison of the above model predictions with our observations of SN~2003du,
to determine whether such an interaction scenario might plausibly reproduce the observed phenomenon.
However, we will not attempt to 
make a detailed comparison between theory of the underlying SN explosion and these observations as has
been done, for example, for SN~1994D and SN~1999by; \citep{h95,h02}.  
Such an analysis would involve fine tuning of the supernova model parameters: the initial 
progenitor, properties of the burning front and the central density of the exploding WD. Instead, we 
have simply used a delayed-detonation model which roughly reproduces the features of a normal Type~Ia SN, 
without attempting to tune this model to match the specific features of SN~2003du.
As an example, in Figure~\ref{obsthe}, the theoretical 
spectra have been plotted at about 3 days before maximum light, along with the May 06 data.
The continuum slope of the model roughly matches the spectrum and the Doppler shifts of lines 
generally agree to within about $ \approx 1000$ km~s$^{-1}$.  (Note that the vertical offset in 
the red is an artifact due to a poor match of the calibration between the two spectroscopic setups.) 
Thus our model, although not tuned in any manner, is a decent match for SN~2003du.

\begin{figure*}
\includegraphics[scale=0.65,angle=270]{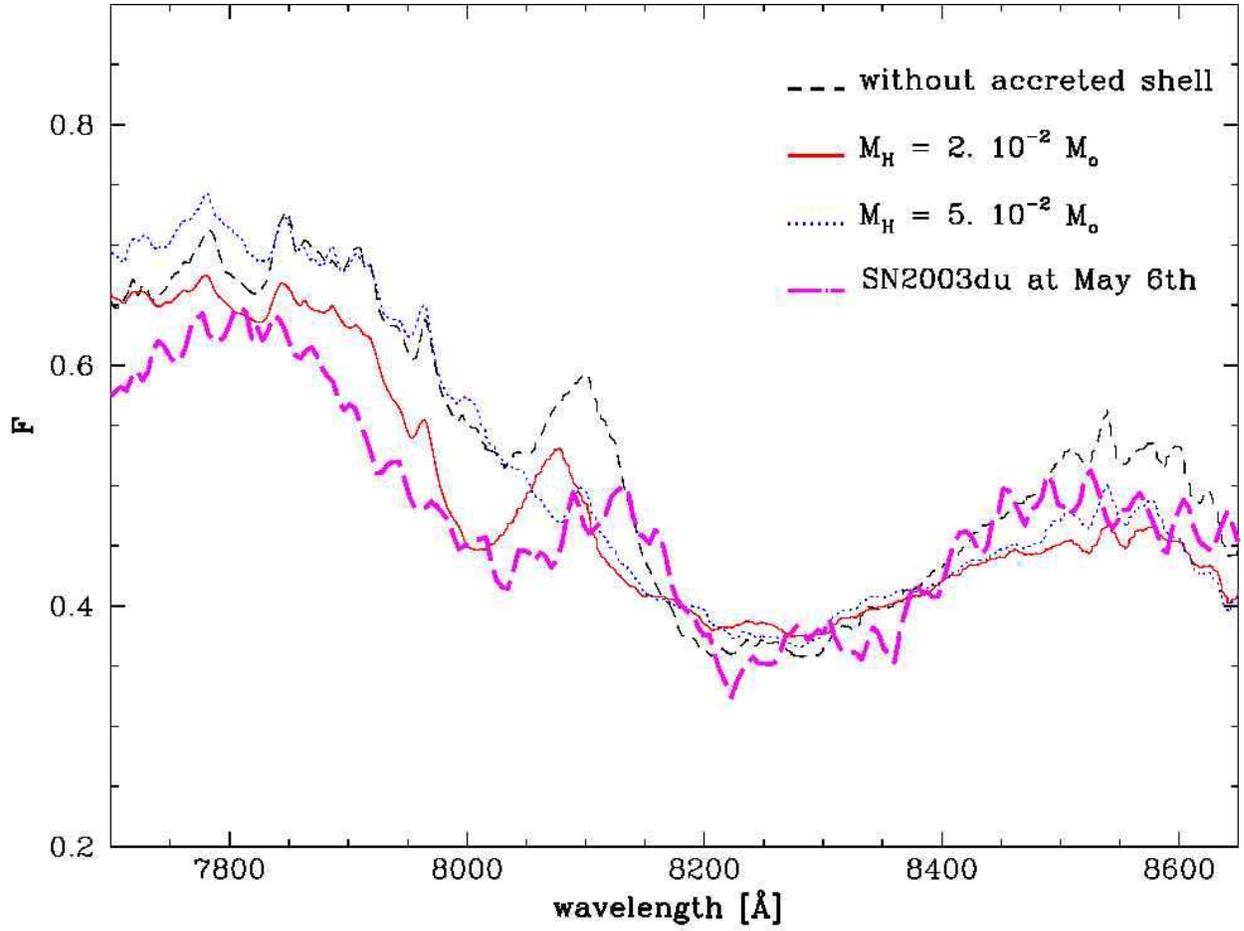}
\caption{\ion{Ca}{2} IR triplet feature observed in SN~2003du on May 6th in comparison with theoretical
models at day 15. The observation is based on the ``red'' spectrum only, and avoids most of the calibration
error seen in Figure \ref{obsthe}. The wiggles in the observations are caused by CCD fringes.}
\label{obsthe_ca}
\end{figure*}

Figure~\ref{obsthe_ca} shows a blow-up of the comparison of Figure~\ref{obsthe}, showing the region of the
\ion{Ca}{2} IR triplet feature.
The strength and Doppler shift of the high-velocity component of the \ion{Ca}{2} is roughly  
consistent with $M_{acc} \approx 2 \times 10^{-2}M_\odot$. The observed line profile is broader than 
the model profile, which is an indication that the absorbing material is somewhat less confined 
than in the model.  As seen in the observations, the model predicts that the high-velocity \ion{Ca}{2} 
starts to become weak around maximum light, and disappears a few days later.  
The higher mass model can clearly be ruled out, as the high-velocity feature in this
model is essentially completely blended with the photospheric absorption, and does not appear as a
well separated feature at all.  

The model without a shell does exhibit a high-velocity component to the \ion{Ca}{2} feature, although the
total absorption is somewhat weaker than the observations (at least with this SN model).  However, this epoch 
(three days before maximum light) was specifically chosen as a ``worst case'' for distinguishing our 
interaction scenario from the short lived ionization effect discussed above. The high-velocity feature 
in the model without a shell 
exhibits an entirely different evolution than is seen in SN~2003du.  As the SN ejecta expands and
cools, the inner and outer regions of the \ion{Ca}{3} region begin to recombine to \ion{Ca}{2}.  
Thus the observed high-velocity feature would shift rapidly to the red, merging with
the photospheric line, and fading over the course of 2 -- 3 days.  

In contrast, the observed evolution of the high-velocity feature more 
closely resembles the expected evolution for a shell that is fixed in velocity space.  The 
line centroid shifts slightly to the red as the feature ages, but the absorption never moves
beyond its upper or lower boundaries.  In the shell model, the slight shift of the  
minimum in the high-velocity component occurs because the inner layers of the shell 
contribute more to the opacity as it expands homologously, and also because the absorption 
depression is formed on the steep blue edge of the main component of \ion{Ca}{2}, which gains 
considerable strength during the observations.  
Note that the observed red 
edge of the high-velocity feature, corresponding to the inner edge of the shell, remains 
constant. This is consistent with the notion that the kinematics of the shell are unchanged, and only the 
relative density of the inner and outer regions of this finite thickness shell are evolving.

Such kinematic stability suggest a shell in free expansion, which implies that most of the circumstellar
matter must have been accumulated prior to the first spectroscopic 
observations.  This, in turn, imposes an  upper limit for the distance between the circumstellar material
and the WD progenitor of $D_{matter} \leq v_{shell}* t \approx 1.5 \times 10^{15}$cm.
Such a distance constrains the origin of the material, which might originate from an accretion 
disk around the white dwarf, mass filling the Roche lobe of the donor star, a common envelope in 
which the WD is embedded, or perhaps a period of very high mass loss immediately prior to the 
explosion.  The last is rather unlikely, as the period of mass loss must 
be shorter than 50 years prior to the explosion, even if we assume a wind of 10 km~s$^{-1}$ which is at 
the low end for red supergiant winds.  Moreover, the interaction with such a wind would convert  
bulk flow kinetic energy at a rate of $2 \times 10^{44}$ erg~s$^{-1}$ (Fig.~\ref{shell}), a 
significant fraction of which should be visible as additional luminosity, but we see no 
evidence for any significant excess in the observed light curve.

A further constraint on the source of the circumstellar material can be inferred from the early light curve.
SN~2003du was discovered on April 22, about 3 days after the explosion in the timeline of our SN
model. At this epoch, SN~2003du was about $2.5^m$ below its maximum light brightness, which is 
typical for normal SNe~Ia (e.g.\ Phillips et al.~ 1999).  A strong interaction with
the surrounding CSM at such an epoch would likely dramatically alter the observed light curve as some fraction
of the kinetic energy of the ejecta is converted into light.  Since the light curve of SN~2003du does not
appear to have been highly unusual, we can infer that the bulk of any extended interaction occurred prior
to the supernova's discovery.  This leads to an estimate for $D_{matter} \leq v_{shell}* t \leq 6 \times 10^{14}cm$.
Furthermore, since the kinetic energy conversion is proportional to the mass accumulation 
rate in the shell, to produce the same shell by interaction with a stellar wind would either result
in very unrealistic luminosities early on, or the deposited energy has to go into expansion work 
rather than luminosity. 

Indeed, the lack of any observed interaction in the light curve argues that the bulk of the 
interaction likely occurred close to the progenitor system, so that adiabatic expansion of the 
freely expanding shell dissipated most of the shock energy.
These constraints tend to argue that the circumstellar material is most 
likely directly related to the progenitor system as in the accretion disk or Roche lobe scenarios, rather
than an effect of the random environment around the progenitor at the time of the explosion.

Note that for small $D_{matter}$, a shell may not stay as confined as assumed in our models
since there will have been more time for internal dynamics to ``smear-out'' the sharp edges of the 
shell.  Such an effect would tend to broaden the shell features, and produce an absorption more akin to
the high-velocity Ca feature observed in SN~2003du.  If the shell
were significantly non-spherical (as in the case of an accretion disk, for example) the kinematics
would also likely be significantly affected.

On the other hand, because of the observed confinement of \ion{Ca}{2} in velocity space, 
the matter cannot be attributed to the outer layers of the WD itself. Redistribution of energy 
during the hydrodynamical phase would produce a smooth velocity profile very similar to the 
freely expanding ejecta (Fig. \ref{model}).  
Since no dense shell would be formed in the outer 
layers, any material from the surface of the WD would have a very low optical depth and the 
observed spectrum would look essentially identical to a normal SN~Ia.

\section{Final Discussion}
{\noindent \bf Summary:}\\
We have presented a series of optical spectra and broad band photometry of SN~2003du
showing that its spectral evolution and light curve resemble that of a normal bright SNe~Ia, except 
for a high-velocity component to the \ion{Ca}{2} IR triplet near 8000\AA. This high-velocity 
component exhibits an expansion velocity of about 18,000 km~s$^{-1}$, significantly larger than the 
photospheric velocity implied from the other spectral features. We propose that this high-velocity Ca feature may be the spectroscopic signature 
of a shell formed by the interaction of the supernova ejecta with the circumstellar environment.

Based on model calculations for the explosion, light curves and spectra, the observable effects
of such a shell have been examined. The model shell is formed by 
interaction between SN ejecta and circumstellar material of solar abundance, with the assumption that the 
circumstellar 
material has been overtaken by the expanding envelope prior to the time of our observations, with 
little or no significant ongoing interaction.  We find that the high-velocity component of the \ion{Ca}{2} IR triplet 
can indeed be understood as being caused by solar abundance 
material piled up in a dense shell behind the shock. This material is likely mixed via 
Rayleigh-Taylor instabilities with the reverse-shocked outer layers of the SN ejecta.
Overall, the SN light curves are little effected except at very early times when the shell is partially
optically thick due to Thomson scattering resulting in larger $(B-V)$ colors by up to $0.3^m$.
Similarly, we find that the optical and IR spectra are otherwize little effected by a shell with an accumulated mass 
of up to a few hundredths of a solar mass. In particular, hydrogen and helium lines are strongly suppressed 
due to the low temperatures in the shells and charge exchange with heavy elements 
being the preferred method of hydrogen recombination.

The main signatures of the shell in our model spectra are the high-velocity component of the \ion{Ca}{2}
IR triplet, and a weaker \ion{O}{1} line near 7300 to 7400 \AA. Although similar in nature, 
these features probe 
different effects.  The Doppler shift of the high-velocity component of \ion{Ca}{2} is a sensitive 
measure of the amount of accumulated matter in the circumstellar shock, whereas the Doppler shift 
of the \ion{O}{1} can be used as a test for mixing of 
shell and envelope material. Unlike the \ion{Ca}{2} feature, the \ion{O}{1} feature is only 
formed with oxygen abundances which exceed the solar value by about 2 
orders of magnitude. 

Related to the interaction process and the deceleration of high-velocity ejecta is the conversion of
kinetic energy.  Assuming reasonable factors of efficiency for the conversion to photons, the 
lack of evidence for any additional associated luminosity 
strongly argues in favor of a very low density environment for SN~2003du
(outside of the immediate vicinity of the progenitor system)
and provides additional constraints for the origin of any circumstellar material.

For SN~2003du, we find that the observations can be understood as a result of a shell formed by the 
accumulation of about $2 \times 10^{-2}M_\odot$ of solar composition material.  
The early light curve data 
strongly suggest that the matter originates from close to or within the progenitor system, as an 
accretion 
disk, the Roche lobe of the companion, or a common envelope. An episode of strong mass loss would 
require unrealistically short durations, very high mass loss rates and probably very high 
luminosities early on. 

{\noindent \bf Relation to Observations of Other SNe~Ia:}\\
Our models indicate that moderate mass shells of around $10^{-2}M_\odot$ show no significant signatures in the
optical and NIR 
except the high-velocity \ion{Ca}{2} feature near 8000 \AA\ and the \ion{O}{1} feature at about 
7300 \AA. In the past, both could easily elude detection because observations often did not extend 
far enough to the red, and even when they did, the increasing noise longwards of 7000 \AA\ and 
contamination by atmospheric absorption bands confuse the issue.
SN~2003du may not, in fact, be all that unusual an event.  Other SNe~Ia (SN~2002cx, SN~2001el)
have shown similar \ion{Ca}{2} features, and a systematic search for this phenomenon in 
SNe~Ia may provide significant constraints on the environment and progenitor systems of these 
objects.

In light of these results, the interpretation of the high-velocity \ion{Ca}{2} IR feature seen 
in SN~2001el and SN~2000cx should be revisited.  Our results here suggest that calcium enrichment may
not be required to reproduce the effect seen in SN~2001el.
The stronger \ion{Ca}{2}
feature and the observed polarization provide somewhat stronger constraints, however, and will
require a more detailed calculation which is beyond the scope of this work.  
We note, however, that the circumstellar phenomena we suggest here for SN2003du are unlike 
those for SN~2003ci which require a large H 
mass at large distances ($\approx 10^{16}$--$10^{17}$ cm; \citealt{wang2003b})

{\noindent \bf Limitations and Future Work:}\\
The analysis presented here, while suggestive, does not provide a definitive conclusion as to the nature of 
the high-velocity Ca absorption, as alternative explanations may still be viable.
In particular, we have not ruled out that the
Ca feature is produced by processed Ca-rich material originating from burning under very special 
conditions in a disk
 or as a consequence of nuclear burning in the WD which causes  ejection of a
high-velocity, Ca-rich filament.
  Rather we have shown that interaction with solar
abundance circumstellar material is a plausible explanation for this phenomenon.  

In examining the
implications for such an interpretation, we have developed observational tools which could be used
to probe the circumstellar environment.  However, these tools still need significant refinement.
Our study has only covered a small 
parameter space and, for the most part, has been limited to the case without ongoing interaction.  
For the formation of 
the shell, we assumed adiabatic shocks.  Depending on the origin of the accreted matter, some of 
the shell material may be accumulated before the phase of homologous expansion has been established
for the SN ejecta, and the shell structure might therefore be modified.  The larger width of the 
high-velocity \ion{Ca}{2} component of SN~2003du (compared to the models) may already be an 
indication of such an effect.  

Also, we have not accounted for the likely 3-D nature of either the SN ejecta or the 
circumstellar material.  Indeed, the polarization measurements for SN~2002el \citep{wang2003a}
indicate that the \ion{Ca}{2} (and thus, in this interpretation, the shell) is not spherical 
but may be toroidal,
or a large scale clump with a significant scale height which selectively blocks light from the 
underlying photosphere and causes polarization \citep{h95,h02,kasen2003}.
Moreover, the polarization confirms that the high-velocity \ion{Ca}{2} in SN~2001el is related to a 
region morphologically separated from the overall SN ejecta. 
The 3-D nature of the absorbing gas will also affect the implied
mass of circumstellar material.  The high-velocity component of the IR \ion{Ca}{2} measures a column
depth along the line of sight 
and the the masses quoted here assume spherical symmetry.  Thus the actual mass may be lower by a factor 
of a few depending on the
covering factor of the absorbing material. 
Finally, asymmetry may also change the early light curve because the luminosity becomes directionally
dependent (H\"oflich 1991), with higher or lower luminosity depending on whether we observe the shell 
from the top or 'equator on' (the latter showing the strong high-velocity Ca II feature). Within our 
configuration, the latter would produce a decrease in the \textit{V} luminosity and even 
redder $(B - V)$ color at early times.
To address these questions, detailed 3D calculations for the interaction of ejecta and environment
should  be performed similar to those for the interaction of the donor star by \citet{mb00} 

\acknowledgments
We would like to thank the staff of the Hobby$^*$Eberly Telescope and McDonald
Observatory for their support.  
We would like to thank E.~Robinson for helpful discussion.
PAH would like to thank David Branch for very useful discussions in Trento.
We would like to acknowledge Carl Akerlof, Don Smith, Eli Rykoff, and 
the members of the ROTSE collaboration for their enormous (and continuing) work on the
design, construction, and implementation of the ROTSE telescopes and software.
The Hobby-Eberly Telescope (HET) is a joint project of the University of
Texas at Austin, the Pennsylvania State University,  Stanford University,
Ludwig-Maximillians-Universitaet Muenchen, and Georg-August-Universitaet
Goettingen.  The HET is named in honor of its principal benefactors,
William P. Hobby and Robert E. Eberly.
  The Marcario Low Resolution 
Spectrograph is a joint project of the Hobby-Eberly Telescope partnership and 
the Instituto de Astronomia de la Universidad Nacional Autonoma de M\'exico.

This research is supported, in part, by NASA grant NAG 5-7937 (PH) and NSF grants
AST0307312(PAH) and AST0098644 (JCW).

\appendix
\section{Details of the SN Model}

\subsection{Numerical Methods}
Our calculations take into account detailed hydrodynamics, nuclear networks,
radiation transport for low and high energy photons, opacities and include solvers to calculate 
the atomic level populations 
in full (albeit simplified) NLTE for both light curves and synthetic spectra utilizing 
accelerated Lambda Iteration and level merging as commonly used in stellar atmospheres
\citep[also known as superlevels, e.g.][]{hilier03,hugeny03,werner03}.  These computational tools have been 
used to carry out several previous analyses of SN~Ia \citep[e.g.][]{h95,h02}. For more details, 
see \citet{hoeflich03a,hoeflich03b}, and references therein.

Spectra are computed using the chemical, density, and luminosity structure as well as the  
$\gamma$-ray deposition resulting from the light curve code. For the detailed atomic models, 
typically between 27 and 137 bound levels are taken into account for the main ionization stages.
For each of these detailed atomic models, neighboring ionization stages have been approximated by 
simplified atomic models restricted to just LTE levels with a few NLTE levels. The energy levels 
and cross sections for bound-bound transitions are taken from \citet{kurucz93} starting at the 
ground state. Here, we use \ion{H}{1}(14/31/91), \ion{He}{1}(15/28/46), \ion{C}{1}
(27/123/242), \ion{C}{2}(23/31/57), \ion{O}{1} (43/129/431), \ion{O}{2}(28/43/75),
 \ion{Mg}{2} (20/60/153), \ion{Si}{2} (35/212/506), \ion{Ca}{2} (41/195/742), \ion{Ti}{2} 
(62/75/592), \ion{Fe}{2} (137/3120/7293), \ion{Co}{2} (84/1355/5396), and 
\ion{Ni}{2} (71/865/3064), where the first, second and third numbers in brackets
denote the number of levels, the number of bound-bound transitions, and the number of discrete 
lines used for the radiation transport. The third number is larger than the second because nearby 
levels within multiplets have been merged for the rates.  In addition, 404508 LTE-lines
are taken into account using an equivalent-two level approach.

For calculating detailed spectra, the explosion models have been remapped from 912 to about 200
radial grid points for the atmospheres with the zones concentrated in the line forming region.
$7.4 \times 10^{4}$ frequency points have been used, oversampling the synthetic
spectra.  In an expanding atmosphere, the frequency and velocity space are coupled and thus
the effective spectral resolution of the synthetic spectra is about 1000.
The noise in the spectra (see Fig.~\ref{model_si}) below this effective resolution is a direct 
result of the oversampling in the frequency space and can be used as an estimate for the internal 
numerical accuracy.

\subsection{SN Model Results}
\begin{figure*}
 \includegraphics[width=7.7cm,angle=270]{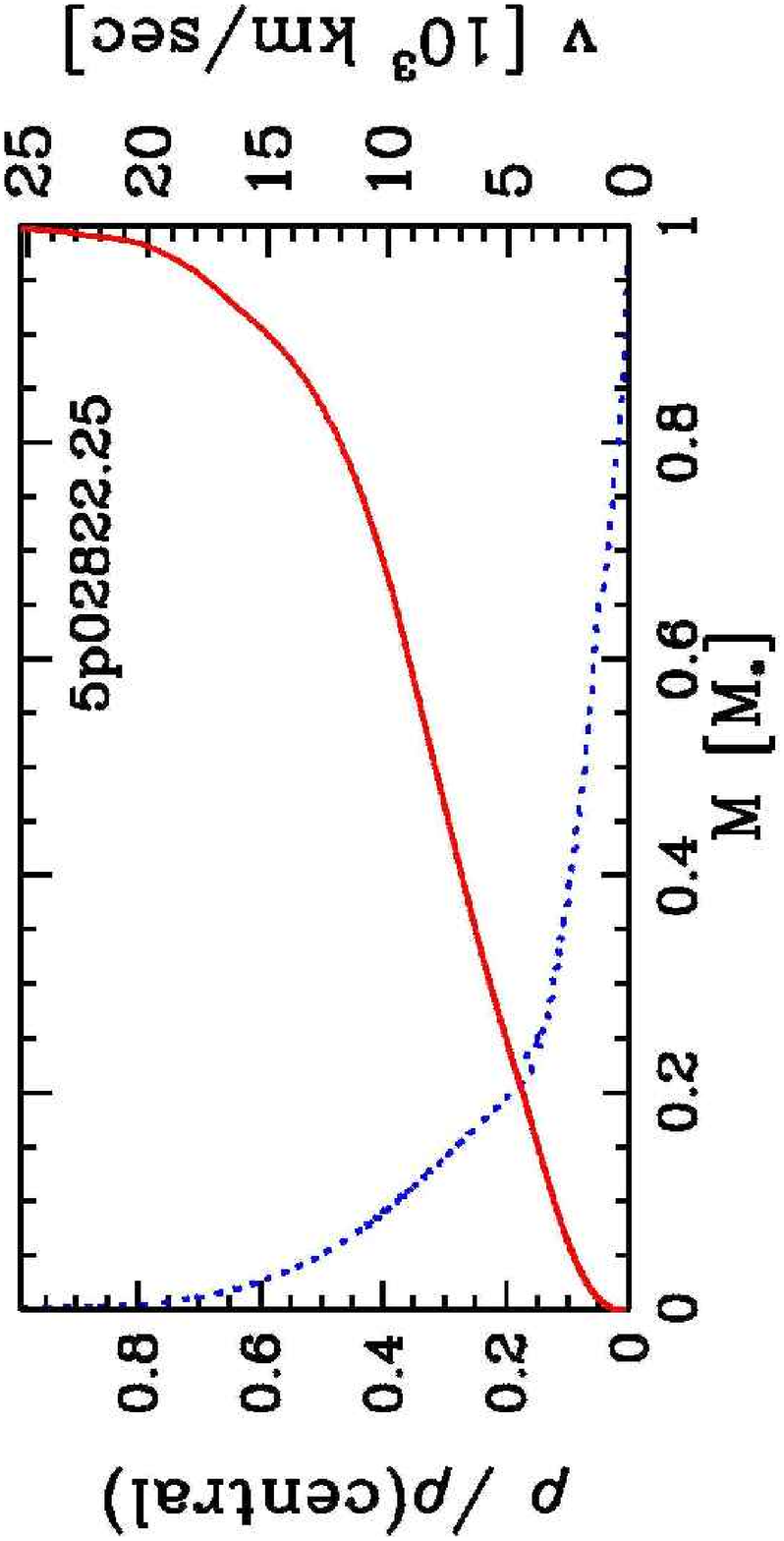}
 \includegraphics[width=6.7cm,angle=270]{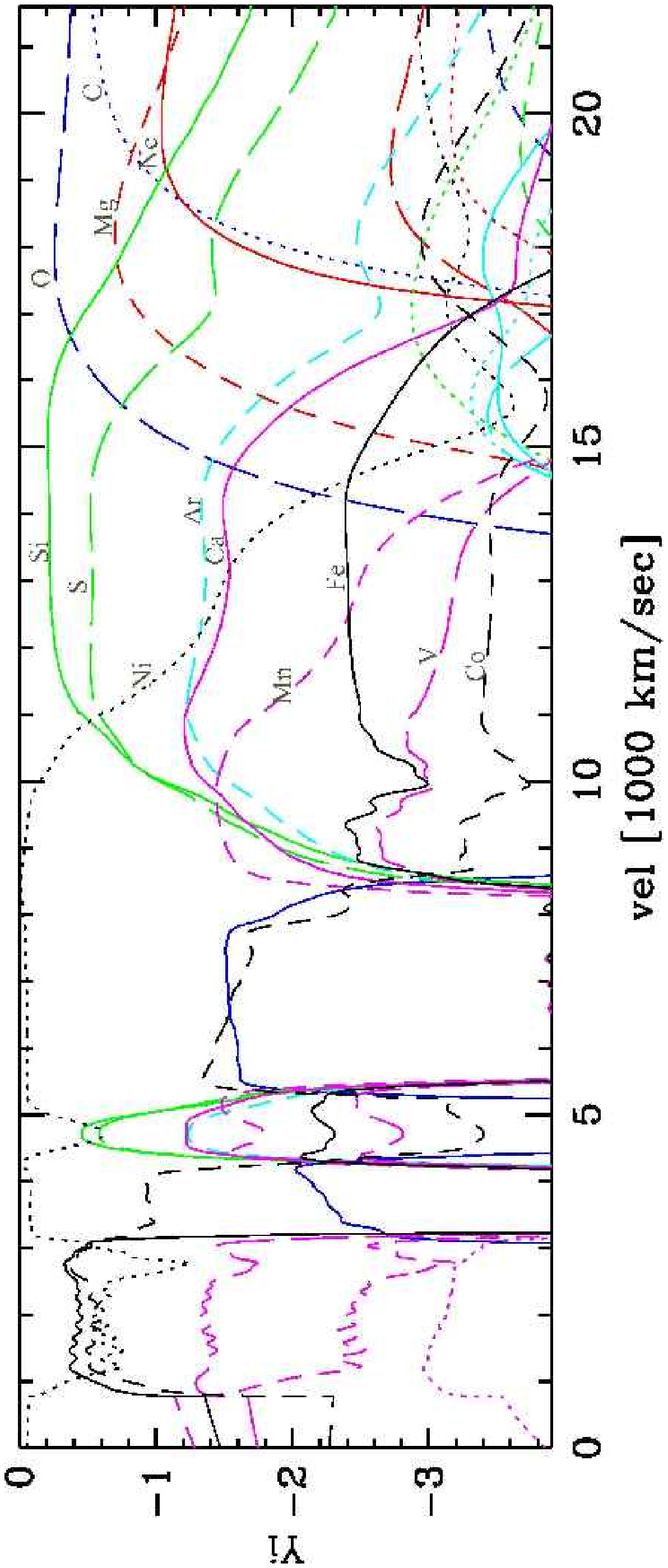}
\caption{Structure of the delayed detonation model. The
density (blue, dotted) and velocity (red, solid) are given as a function of the
expansion velocity during the phase of homologous expansion (upper plot).
The abundances of stable isotopes are given as a function of the expansion velocity
(lower plot). In addition, $^{56}Ni$ is given. The curves with the
highest abundance close to the center correspond to $^{54}Fe$, $^{58}Ni$ and $^{56}Fe$.
 The calculations are based on a delayed detonation model in which a Chandrasekhar 
 mass White Dwarf with central and transition densities of
 $2\times 10^9$ g~cm$^{-3}$ and $2.5 \times 10^7$ g~cm$^{-3}$, respectively.
The progenitor WD has been evolved from a star  with a main sequence mass of
$5 M_\odot$ with solar metallicity.
}
\label{model}
\end{figure*}

The structure of the initial model of the C/O white dwarf is based on a star with 5 solar
masses at the main sequence and solar metallicity which, at the end of its evolution, has lost
all of its H and He-rich layers.  By accretion, its core has grown close to the Chandrasekhar limit.
At the time of the explosion of the WD, the central density is 2.0$\times 10^9$ g~cm$^{-3}$ and 
its mass is close to 1.37$M_\odot$. The deflagration-detonation transition density $\rho_{tr}$ is 
$2.5 \times 10^7$ g~cm$^{-3}$.  During the explosion, about 0.6 
$M_\odot$ of radioactive $^{56}Ni$ are produced. The density and chemical structure
are shown in Figure~\ref{model}. The maximum brightness in \textit{V}, the $(B-V)$ color near 
maximum, the rise time to maximum light, and decline ratio in \textit{V} over 15 days are 
$-19.29^m$, $-0.02^m$, 18.6 days, and $1.0^m$, respectively; typical for normal SNe~Ia. Further 
details for this reference model, including the progenitor evolution, explosion, and light curves 
can be found in \citet{dominguez01} and \citet{h02}, where it is named {\sl 5p0z22.25}.

\clearpage

\begin{deluxetable}{lcccccr}
\tablecaption{SN~2003du Color Evolution\label{colors}}
\tablehead{\colhead{Date} & \colhead{($U-B$)} & 
\colhead{($B-V$)\tablenotemark{a}} & \colhead{($V-R$)} & \colhead{($V-I$)}& 
\colhead{Telescope}}
\startdata
01 May & \nodata & $-0.10$ & $0.01$ & $-0.16$ & HET\\
02 May & \nodata& $-0.02$ & $0.04$ & $-0.07$ & HET\\
06 May & \nodata& $-0.12$ & $-0.08$ & $-0.37$ & HET\\
07 May & $-0.35 \pm 0.06$ & $0.04 \pm 0.05$ & $0.03 \pm -0.05$ & $-0.39 \pm 
0.04$ & 0.8~m\\
08 May & \nodata& $0.00$ & $-0.02$ & $-0.33$ & HET\\
12 May & \nodata& $0.19$ & $0.09$ & $-0.14$ & HET\\
30 May & \nodata& $0.58$ & $0.11$ & $0.06$ & HET\\
\enddata
\tablenotetext{a}{For HET data, {\it B}-band is cut off at 4200 \AA.}
\end{deluxetable}

\clearpage

\clearpage

\clearpage

\clearpage

\clearpage

\clearpage

\clearpage

\clearpage

\clearpage

\clearpage
\clearpage

\end{document}